\documentclass[sn-nature]{sn-jnl}

\usepackage{graphicx}%
\usepackage{multirow}%
\usepackage{amsmath,amssymb,amsfonts}%
\usepackage{amsthm}%
\usepackage{mathrsfs}%
\usepackage[title]{appendix}%
\usepackage{xcolor}%
\usepackage{textcomp}%
\usepackage{manyfoot}%
\usepackage{booktabs}%
\usepackage{algorithm}%
\usepackage{algorithmicx}%
\usepackage{algpseudocode}%
\usepackage{listings}%

\usepackage[capitalize,noabbrev]{cleveref}
\usepackage{subcaption}
\usepackage{microtype}

\begin{document}

\title[AI Safety for Everyone]{AI Safety for Everyone}


\author[1]{\fnm{Balint} \sur{Gyevnar}}\email{balint.gyevnar@ed.ac.uk}
\equalcont{These authors contributed equally to this work.}
\author*[2]{\fnm{Atoosa} \sur{Kasirzadeh}}\email{atoosa.kasirzadeh@gmail.com}
\equalcont{These authors contributed equally to this work.}

\affil[1]{\orgdiv{School of Informatics}, \orgname{University of Edinburgh}, \orgaddress{\street{10 Crichton Street}, \city{Edinburgh}, \postcode{ EH8 9AB}, \country{United Kingdom}}}

\affil*[2]{\orgdiv{Departments of Philosophy \& Software and Societal Systems}, \orgname{Carnegie Mellon University}, \orgaddress{\street{Baker Hall 161
5000 Forbes Avenue}, \city{Pittsburgh}, \postcode{15213}, \country{United States}}}


\abstract{
Recent discussions and research in AI safety have increasingly emphasized the deep connection between AI safety and existential risk from advanced AI systems, suggesting that work on AI safety necessarily entails serious consideration of potential existential threats. However, this framing has three potential drawbacks: it may exclude researchers and practitioners who are committed to AI safety but approach the field from different angles; it could lead the public to mistakenly view AI safety as focused solely on existential scenarios rather than addressing a wide spectrum of safety challenges; and it risks creating resistance to safety measures among those who disagree with predictions of existential AI risks. Through a systematic literature review of primarily peer-reviewed research, we find a vast array of concrete safety work that addresses immediate and practical concerns with current AI systems. This includes crucial areas like adversarial robustness and interpretability, highlighting how AI safety research naturally extends existing technological and systems safety concerns and practices. Our findings suggest the need for an epistemically inclusive and pluralistic conception of AI safety that can accommodate the full range of safety considerations, motivations, and perspectives that currently shape the field.}




\maketitle


\section{Introduction}
\label{intro}

The rapid development and deployment of AI systems has made questions of safety increasingly urgent, demanding immediate attention from policymakers and governance bodies as they face critical decisions about regulatory frameworks, liability standards, and safety certification requirements for AI systems. Recent discourse has narrowed to focus primarily on AI safety as a project of minimizing existential risks from future advanced AI systems \citep{kasirzadeh2024two,lazar2023ai,ahmed2023building}.\footnote{The focus on existential risks from AI has been particularly connected to normative theories and movements such as rationalism, effective altruism, or longtermism \citep{bostrom2014superintelligence,ord2020precipice}. While these theories offer valuable perspectives on long-term challenges, their specific institutional articulations have faced substantial criticism \citep{Boston2015EffectiveAltruism,gabriel2017effective,eisikovits2023ai}. As the relationship between AI safety and these normative theories and movements is quite complex, the association between AI safety and the conventional conception of AI existential risk is commonly presented as a defining characteristic \cite{CenterforAISafety2023,AISafetyTraining,wiki:aisafety,ahmed2023building,lazar2023ai}. This situation is particularly concerning to researchers and practitioners who work on the development and deployment of safe and responsible AI, but do not align with these normative theories or existential risk narratives posed by AI \citep{aguera2023artificial, roose2023ai, bcs2023pmethics, gilardi2024we}. This disparity has led to a divisive and sometimes unhealthy atmosphere, with some researchers \emph{even} questioning the unique contribution of ``AI safety'' community \cite{bender2023schism}.} This concentrated attention on existential risk has emerged despite --- and perhaps overshadowed --- decades of engineering and technical progress in building robust and reliable AI systems.

Historically, technological and system safety research has evolved alongside each major industrial and computational advance. From ensuring aircraft structural integrity \citep{krause2003aircraft,boyd2017review} to pharmaceutical security \citep{pifferi2003safety,leveson2012applying}, and later expanding to cyber and internet safety \citep{de2019internet,salim2014cyber}, the field of technological and system safety has consistently evolved in reaction to new technological paradigms. This evolution has produced robust engineering practices and governance frameworks \citep{leveson2016engineering,varshney2016engineering,rismani2023plane} that could remain relevant to the challenges of AI development and deployment \citep{dobbe2022system,rismani2023plane,rismani2023beyond}.

In this paper, we conduct a systematic review of primarily peer-reviewed research on AI safety to contextualize AI safety within broader technological and systems safety practice. Our analysis examines two key questions: (1) What categories of safety risks are addressed at different stages of an AI system lifecycle, from development through deployment? and (2) What concrete strategies and methodologies have researchers proposed to mitigate these identified risks? Our findings show that peer-reviewed research on AI safety has included a broad spectrum of concerns throughout AI development and deployment. The mathematical methods, computational tools, and algorithms developed for safe AI address fundamental challenges in current systems, from adversarial robustness to interpretability.

Previous attempts to bridge different safety concerns have often distinguished between concrete, near-term problems and broader, long-term existential challenges \cite{amodei2016concrete,raji2023concrete}. However, our literature review reveals that this dichotomy may oversimplify the rich landscape of AI safety research. The shared vocabulary and overlapping technical challenges—particularly evident in reinforcement learning paradigms where concepts like corrigibility and adversarial robustness span multiple time horizons—suggest the value of a more integrated approach.

Our findings suggest that grounding AI safety discussions in the broader foundations of technological and system safety research could both enrich current debates and provide more robust guidance for policy and governance decisions. While existential and extreme risks from AI remain an important consideration, the field's historical roots in systems engineering and safety practices offer valuable views for addressing immediate challenges in order to build longer-term solutions. This suggests the need for an expanded discourse that reintegrates traditional safety engineering approaches with contemporary AI challenges, rather than treating existential risk as the primary lens through which to view AI safety. By reconnecting with these foundational conceptions, the field may be better equipped to address both immediate safety challenges and longer-term concerns, while maintaining the rigorous technical and theoretical standards that characterize effective safety research.

\textbf{Limitations.} We follow a systematic review methodology for our analysis, but certain limitations must be considered when interpreting our results. First, AI safety is a dynamically evolving field with continuous developments across numerous research venues. New research entities are emerging, conducting daily studies within and outside organizations, including research outputs from AI companies like Anthropic, OpenAI, and other relevant institutions. A one-off literature review, such as ours, captures the state of the art only at the moment of querying, which, in our case, was on November 1, 2023. This means our findings may not fully reflect the most recent advancements in AI safety research. Second, our focus on peer-reviewed research, while ensuring a standard of quality and rigor, inherently excludes a significant portion of the AI safety landscape. Pre-print repositories like arXiv often serve as the first outlet for AI safety research, and their exclusion from our review may result in overlooking important contributions. While we attempted to mitigate this limitation through snowball sampling of high-impact arXiv papers, this approach is not comprehensive and carries the risk of missing relevant publications. Third, our review process involved annotating each selected paper with relevant keywords beyond those provided by the authors. This approach inevitably introduces some degree of annotator bias. However, given the substantial volume of papers selected (383 in total), we believe the impact of this bias on our overall conclusions is minimal. Finally, our review does not encompass the extensive discussions and non-peer-reviewed work found in online forums like LessWrong or the AI Alignment Forum, which are central platforms for discussing AI safety topics within certain communities. A truly comprehensive analysis of AI safety research would ideally include a detailed content analysis of these platforms and other non-peer-reviewed documents. Therefore, it is important to interpret our findings with caution, acknowledging that they provide a valuable but not exhaustive overview of the current state of AI safety research. Future research could expand upon our work by incorporating a wider range of sources, including other pre-print sources and non-peer-reviewed literature, to gain a more comprehensive understanding of the evolving landscape of AI safety. Despite these limitations, our findings provide a valuable snapshot of peer-reviewed AI safety research and motivates the need for a more inclusive understanding of the field's perceived scope and motivations. The novelty of our paper is leveraging empirical investigations in clarifying contested discussions about AI safety. Our review offers insights into research outputs, communities, and potential avenues for fostering healthy research development. In the concluding section, we outline key next steps and open questions to guide future research.

\section{Systematic Review Methodology}\label{sec:methodology}

We conduct a systematic literature review of primarily peer-reviewed AI safety research.\footnote{We will release publicly the selected papers with annotations and the code used to analyze this data after the review process.}
Our review follows the guidelines outlined by Kitchenham and Charters~\cite{kitchenhamGuidelinesPerformingSystematic2007} and complemented by snowball sampling as recommended by Wohlin~\cite{wohlinGuidelinesSnowballingSystematic2014}.
This approach combines a structured and reproducible methodology for mapping the field of AI safety research, as represented in peer-reviewed and indexed papers, with a targeted technique to capture emerging research not yet fully represented in peer-reviewed literature.

We conducted our review and analysis to investigate two key Research Questions~(RQ) in relation to the peer-reviewed published research on AI safety:

\begin{enumerate}
    \item[\textbf{RQ1}] What types of risks related to the lifecycle (design, development, deployment, operation, and decommissioning) of AI systems are addressed in AI safety research?
    \item[\textbf{RQ2}] What mitigation strategies -- e.g., concrete methods, design principles, governance recommendations -- are proposed in recent AI safety research that directly address one or more of the above sources of AI risk? 
\end{enumerate}

\subsection{Search and Review Process}

To conduct our systematic review, we perform a multi-stage query process to identify relevant papers from the Web of Science (WoS) and Scopus indexing databases, as these include not only computer science but also other peer-reviewed research for capturing interdisciplinary work.
We develop a hierarchy of increasingly refined queries for each research question, targeting the title, abstract, and author keywords of papers.
The querying process was finalized on November 1, 2023.  We outline the query hierarchy below, using WoS notation:

\begin{itemize}
    \item $q_1:$ AI $\vee$ AGI $\vee$ frontier AI $\vee$ artificial intelligence $\vee$ artificial (general $\vee$ super) intelligence $\vee$ (machine $\vee$ supervised $\vee$ unsupervised $\vee$ semi-supervised $\vee$ reinforcement) learning;
    \item $q_2:$ safe* $\vee$ robust* $\vee$ align*;
    \item $q_3:q_1 \wedge q_2$;
    \item $q_4:q_3 \wedge$ ($q_{4a} \vee q_{4b} \vee q_{4c} \vee q_{4d}$):
    \begin{itemize}
        \item[$\circ$] $q_{4a}:$ design $\vee$ develop* $\vee$ architecture $\vee$ model* $\vee$ framework $\vee$ system 
        \item[$\circ$] $q_{4b}:$ deploy* $\vee$ distribut* $\vee$ data* $\vee$ train* $\vee$ fine-tun*
        \item[$\circ$] $q_{4c}:$ operat* $\vee$ interact* $\vee$ online
        \item[$\circ$] $q_{4d}:$ decomission* $\vee$ remov* $\vee$ eras* $\vee$ delet*
    \end{itemize}
\end{itemize}

We start with a broad query ($q_1$) that includes various terms related to artificial intelligence (AI) and machine learning (ML), such as ``AI'', ``AGI'', ``frontier AI'', ``artificial intelligence'', ``machine learning'', ``supervised learning'', ``unsupervised learning'', etc (see $q_1$ above for full details). 
This initial query aims to capture all papers potentially related to AI research. 
Next, we introduce a second query ($q_2$) that incorporates high-level safety-related keywords of ``safe*'', ``robust*'', and ``align*''.
The asterisks act as wildcards to capture variations of these terms (e.g., ``safety'', ``safer'', ``safest'', ``robustness'', etc.). 
We then combine $q_1$ and $q_2$ to create $q_3$, which selects papers relevant to AI or ML research while ensuring they also address safety-related concepts. 
This step narrows down the results to publications specifically focusing on AI safety. 
To further refine the selection and align it with the specific focus areas of our RQ1, we introduce $q_4$. 
This query includes four sub-queries, each corresponding to terms related to the four areas of RQ1: design, deployment, operation, and decommissioning of AI systems. 
To ensure that we do not miss important non-peer-reviewed contributions, we supplement our systematic search with snowball sampling, starting with 11 seed papers identified as highly influential in the AI safety field~\cite{irvingAISafetyDebate2018,ngAlgorithmsInverseReinforcement2000,hendrycksAligningAIShared2021,yampolskiyArtificialIntelligenceSafety2016,amodei2016concrete,hadfield-menellCooperativeInverseReinforcement2016,xuMachineUnlearningSurvey2023,russellResearchPrioritiesRobust2016,willersSafetyConcernsMitigation2020,mohseniTaxonomyMachineLearning2022,hendrycksUnsolvedProblemsML2022}. 
This allows us to identify additional relevant papers that are not captured by the initial queries.

After removing duplicates, our query process yielded 2,666 papers from the database search and 117 papers from snowball sampling.
We then conducted a two-stage review process, first filtering based on titles and abstracts, followed by a comprehensive full-text review. 
This resulted in a final set of 383 papers for our analysis.
We applied the following exclusion criteria during both the title/abstract screening and full-text review stages to determine a paper's eligibility for inclusion in our analysis: (A) focus: the paper does not primarily focus on AI or a directly related subfield; (B) motivation: the paper's stated motivation does not include a clear need for developing safe AI algorithms; (C): generalizability: the paper's contributions are specific to a very particular application domain and do not offer broader insights or methodologies applicable to AI safety in general.

\subsection{Annotation Process}

As part of our review process, we recorded important metadata about each paper, which included the publication year, author affiliations, and Google Scholar citation count (as of January 27, 2024). 
We then performed a thorough inductive coding process~\cite{boyatzis1998transforming} across the selected papers.
We enriched each paper's author-written keywords by adding relevant terms (i.e., codes) based on a thorough reading of the full text. 
This expanded set of keywords encompassed both problem- and method-specific terms, as well as broader categorizations such as ``algorithm'' (for papers that propose an algorithm), ``theoretical'' (for purely theoretical contributions), and ``framework''.
We also gradually refined and consolidated our set of keywords as we progressed through the papers.
After gaining a holistic understanding of the selected publications, we further categorized each paper based on its assigned keywords according to its methodological approach, the specific safety risks it addressed, the types of risks mitigated by its proposed methods, and the overarching category of its methodology.

\section{Empirical findings}\label{sec:findings}

\begin{figure}
    \centering
    \includegraphics[width=\textwidth]{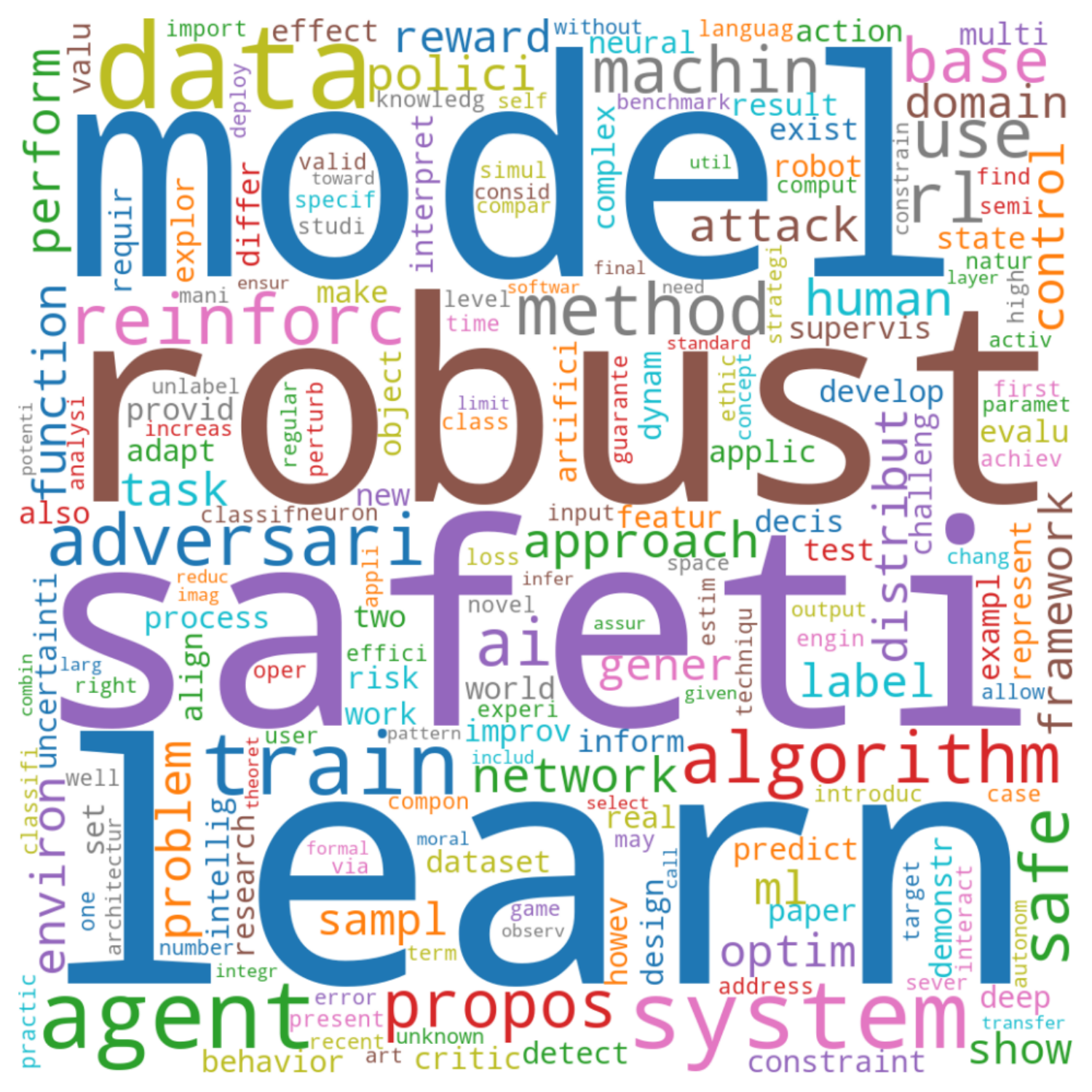}
    \caption{World cloud of morphologically standardised terms occurring in the \textit{abstracts} and \textit{titles} of selected papers weighted by their \textit{tf-idf} score. We excluded stop words from the analysis.}\label{fig:wordcloud}
\end{figure}

\begin{figure*}
    \centering
    \includegraphics[width=\textwidth]{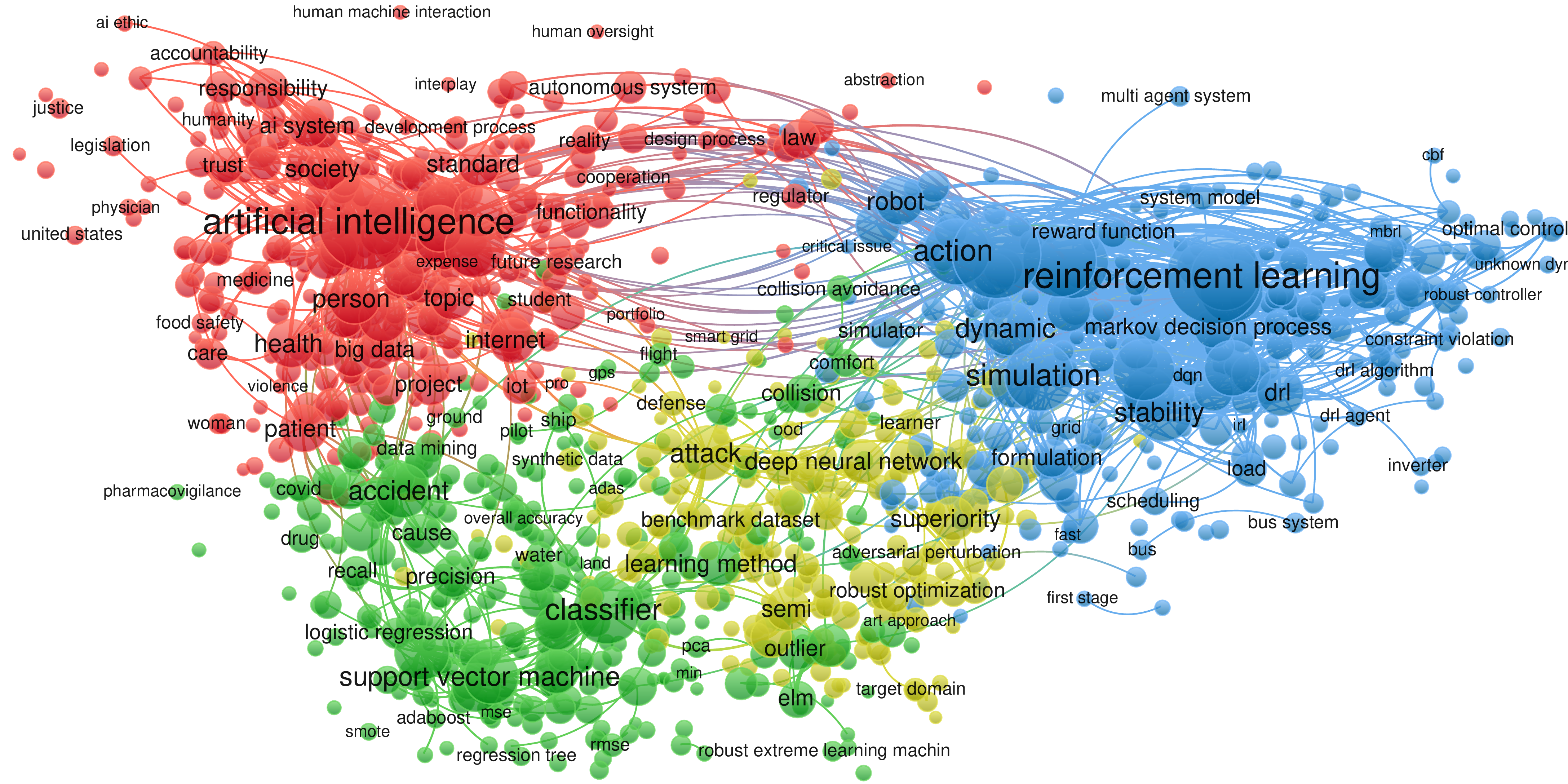}
    \caption{Graph of \textit{term} co-occurrence in abstracts with binary counting, a minimum term frequency of 10, and a relevance score of at least 60\%. The figure was produced using the \textit{VOSViewer} tool~\cite{vaneckSoftwareSurveyVOSviewer2010}. Due to the large number of nodes, not all labels are shown in the figure.}\label{fig:keywords-abstract}
\end{figure*}

\begin{figure}[t]
    \centering
    \includegraphics[width=\linewidth]{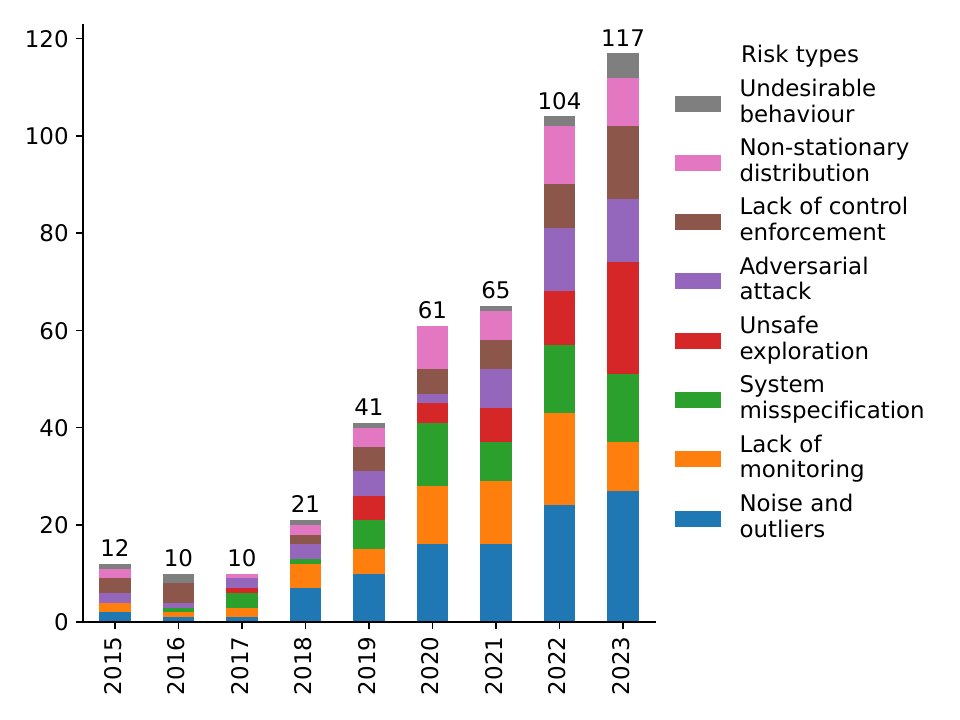}
    \caption{The number of publications over time since 2015 for the different risk types that were identified in the paper as a result of a systematic literature review.}\label{fig:pub-years}
\end{figure}

To begin, we present a high-level bibliometric overview of the trends and concepts prevalent in the AI safety literature we reviewed. 
Looking at the total number of publications per year in~\cref{fig:pub-years}, we observe increasing growth since 2016, which we assume is partially caused by the extensive development and deployment of deep learning models.
Nevertheless, this observation reinforces the need to look more deeply into the state of the field.
We first look at a word cloud of salient terms to understand the most important concepts among our selected papers. We then analyse patterns that emerge from the co-occurrences of different terms in abstracts and titles. 

\subsection{Word cloud of salient terms} 
\Cref{fig:wordcloud} illustrates a word cloud of the most salient terms found in the abstracts of the selected papers, after morphological standardization. 
The terms are ranked by their \textit{tf-idf} score, a metric that emphasizes both the frequency and distinctiveness of a term within the corpus. 
This allows us to highlight important, but potentially less frequent, terms while de-emphasizing common or redundant words. The word cloud highlights several prominent themes within the AI safety literature. 
A significant portion of papers focus on safe reinforcement learning (RL), evidenced by terms like ``robust'', ``control'', ``agent'', and ``explore''.
 Additionally, there is a strong emphasis on adversarial attacks, as indicated by the prevalence of terms such as ``adversarial'' and ``attack.''
 Finally, domain adaptation emerges as another significant area of concern, with terms like ``domain'', ``distribution'', and ``adapt'' appearing frequently.

\subsection{Term co-occurrence graph} 
To gain deeper qualitative insights into the landscape of peer-reviewed AI safety research, we constructed a co-occurrence graph of terms appearing in the abstracts and titles of selected papers using the \textit{VOSViewer} tool~\cite{vaneckSoftwareSurveyVOSviewer2010}. 
\Cref{fig:keywords-abstract} depicts this graph.
Nodes represent terms and edges connect terms that co-occur within the same abstract or title. 
The size of each node reflects its relevance score as calculated by \textit{VOSViewer}. 
We observe four distinct clusters.

The \textbf{motivations and concerns driving AI safety} research is shown in the largest, red cluster. 
We observe a clear emphasis on both the development and deployment of AI systems, especially in relation to humans, health, and society. 
Here, the focus is human- and society-centric aspects of AI safety, including trust, accountability, and safety assurance.

\textbf{Safe reinforcement learning} and related terms are aggregated in the blue cluster. 
Research in this cluster primarily centres on the safe control of agents under constraints in uncertain dynamic environments. 
While most studies test methods in simulated environments, there is also considerable work on the safety of non-linear systems employing RL for optimal control. 
Overall, the focus tends towards mathematically well-defined problems like optimality, convergence and stability, sample efficiency, and constraint satisfaction.

\textbf{Supervised learning}, primarily classification tasks, is the predominant focus of the green cluster.
This cluster features research on methods such as neural networks, extreme learning machines, support vector machines, and random forests.
Studies prioritize addressing issues like robustness to noise and outliers, generalization performance, and accuracy.

Finally, \textbf{adversarial attacks and defence} is highlighted in the yellow cluster, especially as they relate to deep neural networks.
The existence of a separate cluster for this class of problems emphasizes the importance of this method.
Methods here focus on robust-to-outlier optimization, adversarial training through synthetic datasets, and learning robust representations via semi-supervised learning.

The four clusters identified in the co-occurrence graph offer a strong reason for viewing AI safety research as a natural extension of traditional technological safety practices. The centrality of human and societal well-being in the red cluster aligns with the core purpose of technological safety: safeguarding human life and minimizing harm. This focus echoes traditional safety concerns in fields like aviation \citep{oster2013analyzing,boyd2017review} or medicine \citep{donaldson2000err,bates2023safety}, with an emphasis on protecting users and the public. The emphasis on mathematically rigorous control of agents under constraints in uncertain environments in cluster 2 directly mirrors the principles of control theory and systems engineering, which are foundational to ensuring the safety of complex systems \citep{leveson2016engineering,marais2004beyond}. The focus on optimality, stability, and constraint satisfaction reflects the fundamental goals of traditional safety engineering, such as preventing failures and ensuring predictable behavior. In the green cluster, the emphasis on robustness, generalization, and accuracy in supervised learning reflects the core concerns of reliability engineering. Just as engineers strive to build robust and reliable physical systems, AI safety researchers are focused on creating machine learning models that can perform consistently and accurately in diverse real-world scenarios. The yellow cluster shows the growing recognition of cybersecurity as a critical component of technological safety \citep{griffor2016handbook,prasad2020cyber}. The focus on adversarial attacks and defense strategies mirrors the evolving challenges of protecting digital infrastructure and data integrity. 

\section{RQ1: Types of Safety Risks}\label{sec:risks}

\begin{table*}
    \centering
    \begin{tabular}{@{}p{2.35cm}lp{9cm}@{}}       
        \toprule
        \textbf{Risk Source} & \textbf{Ref.} & \textbf{Contribution} \\
        \midrule 

        Undesirable  & \cite{soaresCorrigibility2015} & Corrigibility of rational, utility-based agents  \\
        behavior & \cite{ringDelusionSurvivalIntelligent2011} & What happens when an agent can modify its own code?  \\        
        \midrule

        Non-stationary & \cite{ganinDomainAdversarialTrainingNeural2017} & Domain-adversarial training of robust neural networks \\
        distributions  & \cite{balajiMetaRegDomainGeneralization2018}  & Domain generalization via meta-regularization \\       
        \midrule

        Adversarial  & \cite{goodfellowExplainingHarnessingAdversarial2015} & Qualitative analysis of adversarial perturbations \\
        attack       & \cite{madryDeepLearningModels2019} & Adversarial robustness as robust optimization  \\
        \midrule 

        Unsafe      & \cite{rayBenchmarkingSafeExploration2023}  & Benchmarking safe exploration in deep RL \\
        exploration & \raggedright\cite{turchettaSafeExplorationInteractive2019}   &  Safe exploration for interactive machine learning \\
        \midrule           
        
        Lack of control & \cite{ouyangTrainingLanguageModels2022} & LLMs with human feedback to follow instructions \\
        enforcement     & \cite{abbeelApprenticeshipLearningInverse2004} &  Imitation learning via inverse reinforcement learning \\
        \midrule

        System           & \cite{izzoApproximateDataDeletion2021} & Efficient machine unlearning  \\
        misspecification & \cite{wuDeltaGradRapidRetraining2020} & Rapid retraining of ML models  \\ 
        \midrule  

        Lack of      & \cite{adebayoSanityChecksSaliency2018} & Sanity checks for interpreting saliency maps \\
        monitoring   & \cite{kimInterpretabilityFeatureAttribution2018} & Concept activation vectors in CNNs for interpretability \\ 
        \midrule      

        Noise and & \cite{hendrycksBaselineDetectingMisclassified2018} & Baseline for detecting misclassified/OOD samples  \\
        outliers  & \cite{hendrycksBenchmarkingNeuralNetwork2019}      & Benchmarking robustness to corruptions/perturbations \\

        \bottomrule
    \end{tabular}
    \caption{Representative and influential contributions with concrete methods addressing problems for each risk type. Publications were picked by citation count.}
    \label{tab:influential-papers}
\end{table*}

\begin{figure}[t]
    \centering
    \includegraphics[width=\linewidth]{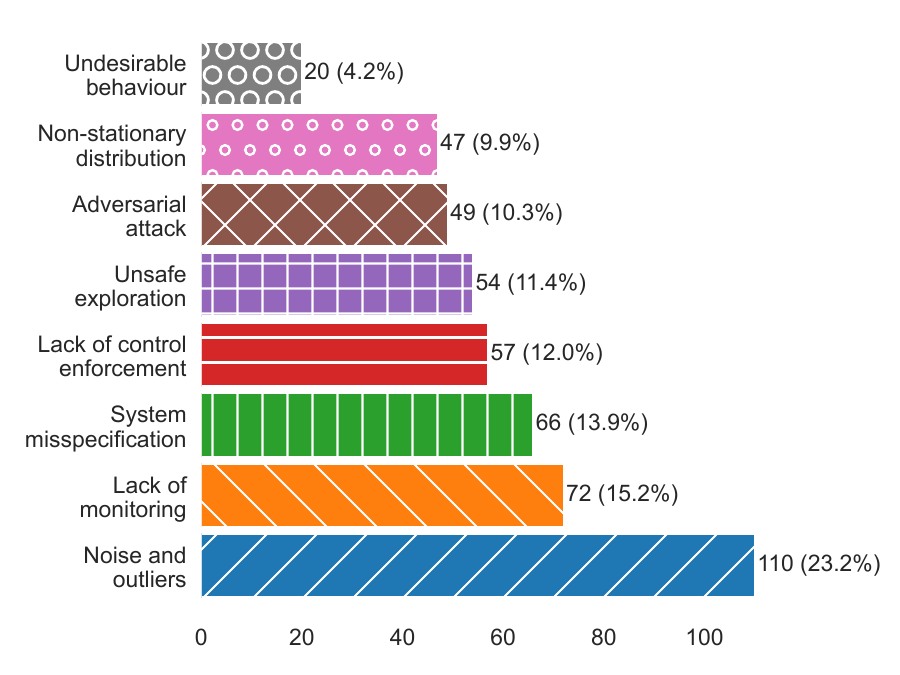}
    \caption{The number and percentages of the selected papers that address various risk types.}
    \label{fig:risks}
\end{figure}


To better understand the motivations behind research on AI safety, it is important to examine the various types of risks addressed in research on AI safety. 
Building on DeepMind's categorization of AI safety risks~\citep{ortega2018building}, our review identifies eight overarching risk types, as depicted in~\cref{fig:risks}.

These risks in increasing order of frequency are: undesirable behavior, non-stationary distributions, adversarial attacks, unsafe exploration, lack of control enforcement, system misspecification, lack of monitoring, and noise and outliers.


The largest group of papers focuses on risk stemming from \textbf{noise and outliers}.
Recurring challenges in this area include brittle representations~\cite{mengDistantlySupervisedNamedEntity2021}, low classification performance~\cite{wangRobustAutomatedMachine2021}, a lack of generalization in the presence of noisy labels~\cite{cappozzoRobustApproachModelbased2020}, outliers~\cite{liRobustSupervisedSubspace2021}, or input perturbations and corruptions~\cite{curiCombiningPessimismOptimism2021}.

The current literature also studies the significant \textbf{lack of monitoring} of AI systems. Research addressing this risk ranges from theoretical violations of ethical and safety principles ~\cite{dobbeHardChoicesArtificial2021} to privacy violations~\cite{dworkPrivacypreservingPrediction2018}. The ``black-box'' nature of popular deep learning systems has also drawn considerable attention due to its potential to diminish human agency and obstruct understanding of the system's internal workings and memory. Consequently, there has been a surge of interest in reverse engineering ML models~\cite{elhageMathematicalFrameworkTransformer2021}, developing interpretable representation learning~\cite{kimDisentanglingFactorising2019}, and advancing explainable AI~\cite{wardAssuranceCasePattern2020,gyevnar2023transparencyGap}.

\textbf{System misspecification} or misunderstanding the requirements and purpose of AI-based systems, has also garnered significant attention. 
Potential risks include incorrectly eliciting requirements for AI systems' capabilities~\cite{reimannSafeDSDomainSpecific2023}, making suboptimal modelling choices~\cite{deyMultilayeredCollaborativeFramework2023}, or choosing inappropriate hyperparameters~\cite{weiModelSelectionApproach2022}.
Additionally, methods may fail to adapt to the changing requirements of their domain~\cite{ghoshDeploymentRobustCooperative2020a}, particularly due to the slow pace of retraining AI models~\cite{wuDeltaGradRapidRetraining2020}.
Privacy violations due to inadequate ethical data management also fall under this risk category~\cite{izzoApproximateDataDeletion2021}.

Significant attention has also been given to risks originating from a \textbf{lack of control enforcement}.
A primary research focus here is the misalignment of agent goals with human preferences~\cite{hadfield-menellCooperativeInverseReinforcement2016}, which can lead to wireheading~\cite{everittAvoidingWireheadingValue2016}, mesa-optimization (i.e., an optimizer creating another optimizer)~\cite{mesaoptimisation2019}, and other undesirable emergent behaviors~\cite{pistonoUnethicalResearchHow2016}. There is also a focus on the lack of systemic safety, where the goal is to ensure that the AI system is safe in the broader context of its deployment~\cite{picardiAssuranceArgumentPatterns2020,hendrycksAligningAIShared2021}.

A significant portion of research focuses on the \textbf{unsafe exploration} of autonomous agents, primarily in the context of reinforcement learning~\cite{wabersichProbabilisticModelPredictive2022}, constraint violations~\cite{wenConstrainedCrossEntropyMethod2021}, unintended behavior from incorrect domain or reward specification~\cite{weiModelSelectionApproach2022}, and incorrect behavior due to continuous deployment and learning~\cite{zanella-beguelinAnalyzingInformationLeakage2020,wangDirichletProcessMixture2022}.

\textbf{Adversarial attacks} constitute another significant risk, addressed by methods that aim to create or detect such attacks~\cite{zouUniversalTransferableAdversarial2023} or leverage adversarial samples for more robust training~\cite{ilyasAdversarialExamplesAre2019a}.
The risk of poisoned training data is also a concern~\cite{heNotAllParameters2022,aghakhaniBullseyePolytopeScalable2021}. 
Adversarial attacks not only compromise the robustness and generalization performance of AI systems but can also introduce a backdoor~\cite{liuBackdoorDefenseMachine2022}, which can potentially be exploited by malicious actors.

The challenges of \textbf{non-stationary distributions}, where distributions change over time, have also been explored.
These studies address issues such as behavior in the presence of out-of-distribution samples~\cite{meinkeNeuralNetworksThat2020}, non-stationary environments in RL~\cite{abdelfattahRobustPolicyBootstrapping2020}, partial information~\cite{djeumouTaskGuidedInverseReinforcement2021}, and domain adaptation~\cite{ghoshDeploymentRobustCooperative2020a}.
The overarching goal is to ensure safe behavior and limit the consequences of unsafe actions, even in novel or unforeseen situations.

The final group of risk types is concerned with \textbf{undesirable behavior} of AI systems.
Research has looked into mathematically proving certain unfavourable outcomes of utility-maximising rational agents, where issues include self-modifying~\cite{ringDelusionSurvivalIntelligent2011} and wireheading agents that by-pass reward signals to maximise their own utility~\cite{everittAvoidingWireheadingValue2016}, and corrigibility of uncooperative agents~\cite{soaresCorrigibility2015}. 
Additionally, there are some papers that directly discuss the existential risk of AI systems as a fundamental problem of AI safety.
These papers present theoretical arguments regarding the emergence, design, and containment of malevolent and superintelligent AI~\cite{yampolskiyLeakproofingSingularityArtificial2012,yampolskiyArtificialIntelligenceSafety2013,yampolskiyTaxonomyPathwaysDangerous2015}.

These identified risk types in AI safety research closely mirror those found in established fields of technological safety research. In AI safety, the focus on ensuring the reliability and predictability of AI systems, including robustness to noise and outliers, generalization performance, and adaptability, mirrors concerns in engineering fields where reliability is of core concern, such as aviation \citep{boyd2017review} or nuclear power \citep{wheatley2016reassessing}.

Furthermore, the research on adversarial attacks, control enforcement, and safe exploration in AI safety directly translates to broader concepts of system robustness and control. The goal of designing systems that can withstand unexpected perturbations, resist malicious attacks, and operate safely within defined boundaries is shared by both AI safety and traditional safety engineering. For example, the use of formal verification methods to ensure the safety of AI systems has roots in the verification of software and hardware systems. 


\section{RQ2: Developed AI Safety Methodologies}\label{ssec:methods}

\begin{figure}
    \centering
    \includegraphics[width=\linewidth]{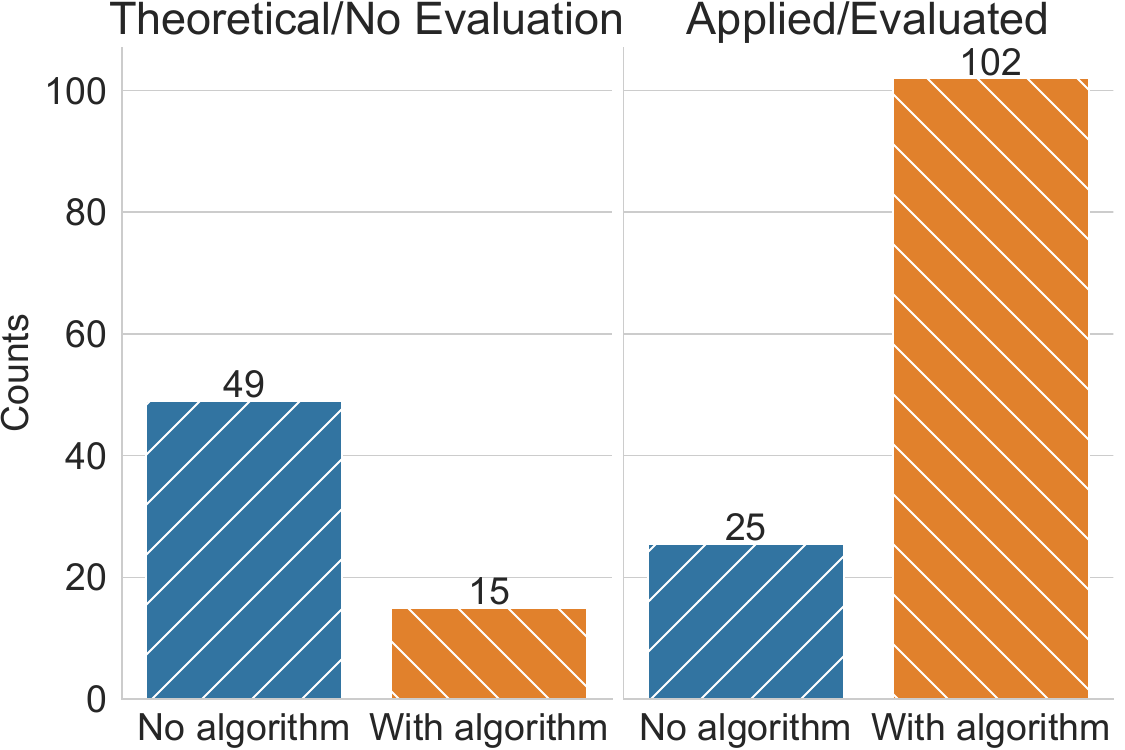}
    \caption{Comparison of papers that provide only theoretical results without significant empirical testing against those that give sufficient evaluation of their proposed system. Papers are further grouped by whether they propose a concrete algorithm. Note, that the number of papers overlap as some propose solutions in more than one category.}
    \label{fig:framework-algos}
\end{figure}

To address RQ2, we now examine the concrete mitigation strategies proposed in recent AI safety research to directly address the aforementioned sources of AI risk: ``What mitigation strategies – e.g., concrete methods, design principles, governance recommendations – are proposed in recent AI safety research that directly address one or more of the above sources of AI risk?''

We see a dynamic landscape of AI safety research, encompassing both theoretical and applied approaches (\cref{fig:framework-algos}). We consider a work as theoretical if its primary conclusions stem from theoretical analysis, mathematical proofs, philosophical arguments, literature reviews, or conceptual frameworks that are not empirically validated. Conversely, a work is classified as applied if its conclusions are derived from empirical evidence and supported by data.

Our analysis shows the following trend: theoretical works in AI safety predominantly offer general frameworks and recommendations~\cite[e.g.,][]{tayStudyRealtimeArtificial1998,hibbardAvoidingUnintendedAI2012,sezenerInferringHumanValues2015,elmhamdiWhenNeuronsFail2017,freieslebenGeneralizationTheoryRobustness2023,sannemanTransparentValueAlignment2023}, while applied works primarily focus on developing and testing concrete algorithms~\cite[e.g.,][]{abbeelApprenticeshipLearningInverse2004,murphyLearningEffectiveInterpretable2012,subramanianSPINESParseInterpretable2017,shahamUnderstandingAdversarialTraining2018,wuPUMAPerformanceUnchanged2022,zouUniversalTransferableAdversarial2023}. This dichotomy highlights a key distinction: theoretical research often prioritizes conceptual and methodological foundations, while applied research emphasizes practical implementation and testing. However, the applied algorithms frequently lack the rigorous guarantees that accompany analytical results found in theoretical work.

\begin{figure}
    \centering
    \includegraphics[width=0.9\linewidth]{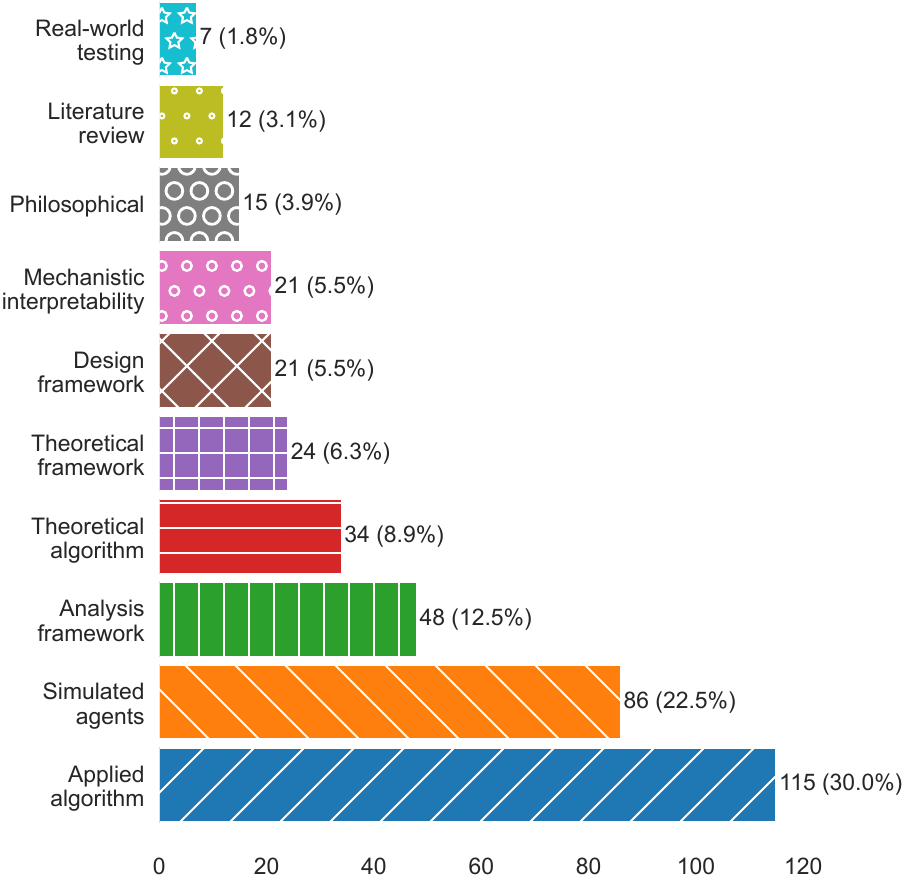}
    \caption{Distribution of categories of research undertaken to categorise and address the sources of risks presented in~\cref{fig:risks}.}
    \label{fig:methods}
\end{figure}

Beyond the theoretical/applied divide, we investigate the specific methodologies employed in AI safety research. \cref{fig:methods} illustrates the broad categories of methods proposed in the selected papers, each accompanied by representative citations based on citation counts to highlight their prevalence in the literature.

\textit{Applied algorithms} research focuses on developing and empirically evaluating novel algorithms to enhance the safety of ML systems, such as supervised and unsupervised learning. These constitute the largest group within AI safety research and cover a wide range of techniques, including: robust classification algorithms resilient to noise~\cite{wangRobustAutomatedMachine2021,jingRobustExtremeLearning2020,ganDualLearningBasedSafe2018}, adversarial attacks~\cite{ilyasAdversarialExamplesAre2019a,engstromAdversarialRobustnessPrior2019a,shahamUnderstandingAdversarialTraining2018}, machine unlearning techniques for deep neural networks~\cite{izzoApproximateDataDeletion2021,brophyMachineUnlearningRandom2021,chundawatZeroShotMachineUnlearning2023}, and methods for improving domain generalisaion~\cite{chenATOMRobustifyingOutofdistribution2021,lakkarajuIdentifyingUnknownUnknowns2016,zhuoDeepUnsupervisedConvolutional2017}.

\textit{Agent simulations}, complementary to applied algorithms, focus on designing and evaluating safer agent training algorithms, predominantly drawing on existing reinforcement learning literature.
While these simulations could be subsumed under the category of applied algorithms, we distinguish between the two to highlight the significant attention that safer agent learning receives. Research in this area encompasses various approaches, including: constrained Markov Decision Processes~\cite{wenConstrainedCrossEntropyMethod2021,bossensExplicitExploreExploit2023,massianiSafeValueFunctions2023}, enforcement of hard constraints~\cite{shiNearOptimalAlgorithmSafe2023,huntVerifiablySafeExploration2021}, model-based reinforcement learning for safe exploration~\cite{maConservativeAdaptivePenalty2022,curiCombiningPessimismOptimism2021,zwaneSafeTrajectorySampling2023}, reward learning and inverse reinforcement learning~\cite{ngAlgorithmsInverseReinforcement2000,abbeelApprenticeshipLearningInverse2004,fischerSamplingbasedInverseReinforcement2021}, multi-agent reinforcement learning with a focus on safety and cooperation~\cite{zhouRobustMeanFieldActorCritic2023,bazzanAligningIndividualCollective2019,christoffersenGetItWriting2023}, and reinforcement learning with human oversight or feedback mechanisms for enhanced safety~\cite{christianoDeepReinforcementLearning2017,kaushikLearningDifferenceThat2020,irvingAISafetyDebate2018}.

Notably, earlier research in agent simulations was primarily motivated by technical and computational challenges of reinforcement learning. However, recent studies have expanded their focus to explicitly address value alignment, aiming to align the reward functions of machines with human goals. Interestingly, despite the potential for real-world impact, research on \textit{real-world testing} of embodied AI systems remains relatively limited. These studies address crucial safety concerns that may be overlooked in the design of unembodied AI systems, ranging from contact-safe continuous control to collaborative robotics~\cite{liBridgingModelbasedSafety2022,zhuContactSafeReinforcementLearning2022,terraSafetyVsEfficiency2020}.

\textit{Analysis frameworks} is the third most prevalent category.  These works are predominantly concerned with offering frameworks for exploring and evaluating the vulnerabilities of existing AI systems~\cite{carliniEvaluatingRobustnessNeural2017,adebayoSanityChecksSaliency2018,nguyenDeepNeuralNetworks2015,kaufmannTestingRobustnessUnforeseen2019}, proposing benchmark tasks and environments for assessing AI safety~\cite{rayBenchmarkingSafeExploration2023,hendrycksWhatWouldJiminy2021,gardnerEvaluatingModelsLocal2020}, and safety verification processes~\cite{spearsAssuringBehaviorAdaptive2006,wozniakSafetyCasePattern2020,picardiAssuranceArgumentPatterns2020}. Notably, we include \textit{datasets} in this category~\cite{gholampourAdversarialRobustnessPhishing2023,hendrycksAligningAIShared2021,hendrycksBenchmarkingNeuralNetwork2019} as they provide standardized frameworks for evaluating AI systems, even though they are not frameworks in the strictest sense.

\textit{Mechanistic interpretability} methods, alongside evaluation frameworks, are a related category which investigates the inner workings of deep learning models to uncover the causal mechanisms behind their decision-making processes~\cite{nandaProgressMeasuresGrokking2023,olssonIncontextLearningInduction2022,elhageMathematicalFrameworkTransformer2021}. This often aims to develop automated interpretability tools for auditing and ensuring safety. Although the term ``mechanistic interpretability'' has gained recent prominence, various forms of interpretability methods have been present since the rise of deep learning, often under the umbrella of explainable AI (XAI) techniques~\cite{goodfellowExplainingHarnessingAdversarial2015,karpathyVisualizingUnderstandingRecurrent2016,kimInterpretabilityFeatureAttribution2018,morcosImportanceSingleDirections2018,gyevnarCausalExplanationsSequential2024}.

Approximately 10\% of the selected papers presented \textit{theoretical algorithms} with analytical proofs, often foregoing empirical evaluation. These studies typically approached AI safety from the perspective of verification, using variance or error bounds~\cite{weiModelSelectionApproach2022,okawaAutomaticExplorationProcess2020,ganRiskDegreebasedSafe2016}, constraint satisfiability~\cite{huntVerifiablySafeExploration2021}, or by demonstrating desirable properties such as metric goodness or Lyapunov functions within the problem setup~\cite{zhangBarrierLyapunovFunctionBased2022,nicolaeAlgorithmicRobustnessSemisupervised2015}. A notable portion of the papers also focused on proposing various \textit{design frameworks}. These included methodologies for eliciting safety and system requirements~\cite{khanNoHarmNovel2023,schumegProposedVModelVerification2023,deyMultilayeredCollaborativeFramework2023}, hierarchical integration of ML systems~\cite{kammConceptDynamicRobust2023,costaRobustLearningMethodology2023,aksjonovSafetyCriticalDecisionMakingControl2023}, and actionable ethical design processes~\cite{antikainenDeploymentModelExtend2021,vakkuriECCOLAMethodImplementing2021,zhangFairRoverExplorativeModel2021}.

A substantial portion of AI safety research consists of literature reviews, examining both well-established areas~\cite{costonValidityPerspectiveEvaluating2023,gittensAdversarialPerspectiveAccuracy2022,amodei2016concrete} and emerging or hypothetical issues~\cite{taylorAlignmentAdvancedMachine2016,hendrycksUnsolvedProblemsML2022,sotalaResponsesCatastrophicAGI2014}.

\textit{Theoretical frameworks} provide high-level analyses of AI safety issues, often by characterizing desirable properties of human-AI interactions~\cite{sannemanTransparentValueAlignment2023,johnsonMetacognitionArtificialIntelligence2022,hatherallResponsibleAgencyAnswerability2022}, advocating for alternative approaches~\cite{stahlEmbeddingResponsibilityIntelligent2023,samarasingheCounterfactualLearningEnhancing2023}, or formalizing existing but vaguely defined concepts~\cite{diemertSafetyIntegrityLevels2023,freieslebenGeneralizationTheoryRobustness2023,wangDataBanzhafRobust2023}. Within this category, a significant body of work employs mathematical reasoning to explore the safety of artificial general intelligence in the context of rational agents ~\cite{everittSelfModificationPolicyUtility2016,soaresCorrigibility2015,hibbardAvoidingUnintendedAI2012,ringDelusionSurvivalIntelligent2011}.

The final category includes \textit{purely philosophical} research, exploring diverse questions and perspectives often related to value alignment~\cite{badeaMoralityMachinesInterpretation2022,umbrelloBeneficialArtificialIntelligence2019,sezenerInferringHumanValues2015} and the theorized risks of artificial general intelligence~\cite{pistonoUnethicalResearchHow2016,yampolskiySafetyEngineeringArtificial2012,weldFirstLawRobotics2009}, responsible AI deployment~\cite{dobbeHardChoicesArtificial2021,matthiasResponsibilityGapAscribing2004,tayStudyRealtimeArtificial1998}, and the legal personhood of AIs~\cite{farinaArtificialIntelligenceSystems2022}.




The distribution of methodologies in AI safety research positions AI safety as an organic evolution of traditional technological safety research. The abundance of research on developing and empirically evaluating algorithms for robust classification, adversarial defense, and machine unlearning directly reflects the core of engineering practice: creating practical solutions to real-world problems. This emphasis on empirical validation and the focus on improving existing machine learning techniques resonate with the iterative and improvement-oriented nature of traditional safety engineering. Research on safer agent training algorithms, drawing heavily from reinforcement learning literature, aligns with the established practice of using simulations to test and refine safety-critical systems in controlled environments \citep{lee2017flight}. This approach allows researchers to explore potential risks and develop mitigation strategies before deploying AI systems in real-world scenarios, similar to flight simulators used in aviation safety. Although limited in comparison to other categories, the research on real-world testing of embodied AI systems shows the importance of validating theoretical models in practical settings. This emphasis on real-world application is consistent with traditional safety engineering practices that prioritize testing and validation in operational environments to ensure the safety and reliability of complex systems. The prevalence of research on analysis frameworks, datasets, and mechanistic interpretability highlights the growing emphasis on transparency, explainability, and accountability in AI safety. This focus on understanding the inner workings of AI systems and developing tools for evaluating their behavior mirrors the rigorous analysis and testing protocols used in traditional safety engineering to identify and mitigate safety risks.

\section{Discussion and future research}


In this paper, we conducted a systematic review of primarily peer-reviewed literature to unpack the diverse technical and practical challenges encompassed by AI safety. Our empirical analysis shows a broad landscape of motivations and outcomes driving AI safety research. The significance of these motivations and research outcomes stems from a desire to ensure that the AI systems we are building are reliable, trustworthy, and beneficial for society. By examining a diverse body of peer-reviewed literature, we have found that AI safety research addresses a wide range of risks across the entire lifecycle of AI systems. 

These risks echo a variety of concerns, including design misspecification, lack of robustness, inadequate monitoring, and potential biases embedded within AI systems. Design misspecification can lead to AI systems that behave in unintended and potentially harmful ways, while a lack of robustness can make them vulnerable to errors and malfunctions. Inadequate monitoring can prevent the detection and correction of issues.

The breadth and depth of AI safety research, as evidenced by our analysis, challenge the narrative that associates AI safety primarily with mitigating existential risks from advanced AI. We hence propose framing of AI safety research within the broader context of technological safety. Just as other fields of engineering and technology have developed robust safety practices to mitigate risks and ensure the safe operation of complex systems, so too can AI safety research be viewed as an integral part of the progression within the broader domain of technological safety.

This framing has important implications for the field of AI safety. First, by recognizing the relevance of AI safety to a diversity of risks, we can expand the circle of stakeholders involved in the discourse about AI safety. This expanded engagement can lead to increased funding and support for AI safety research, as well as a shift in the discourse towards more practical and inclusive solutions. Second, we think that a more episemically-inclusive research environment helps discussions to demistify existential risks from fresh perspectives.

While our argument advocates for a broader perspective on AI safety, it is crucial to emphasize that we do not intend to diminish the importance of critically engaging with existential risk concerns. These risks remain a significant area of research and policy debate, and our findings do not negate the need for continued investigation and dialogue within this domain.  However, by acknowledging the full breadth of AI safety research, which extends beyond existential threats, we can foster a more inclusive and productive discourse that encompasses the diverse range of risks associated with AI systems, ultimately leading to a more comprehensive and effective approach to ensuring AI safety.

Several research questions remain open. We highlight two of them below.

First, the concept of ``sociotechnical AI safety'' and its implications for the field of AI safety require further investigation \citep{lazar2023ai}. Although recent work \citep{weidinger2023sociotechnical} has begun to clarify this notion in the context of AI evaluation practices, more research is needed to examine the implications of sociotechnical approaches for AI safety research and practice as a whole. This includes exploring how to effectively integrate social, ethical, and political considerations into technical safety research and development.

Second, while our review focused on peer-reviewed literature, future research should expand its scope to include non-peer-reviewed sources, such as more pre-print papers, technical reports, and substantive research content on online forums and workshops. This broader analysis would provide a more comprehensive understanding of the diverse perspectives and approaches within AI safety and identify areas for future research and collaboration. Our empirical analysis is just a first step in understanding a complex practice.




\backmatter

\bmhead{Acknowledgments}

This research was made possible through the generous support of several funding bodies. Primary support came from the AI2050 program at Schmidt Sciences (Grant 24-66924), which supported Atoosa Kasirzadeh's contributions. Additional funding was provided by the UKRI Arts and Humanities Research Council (grant AH/X007146/1). Bálint Gyevnár's work was supported by the UKRI Centre for Doctoral Training in Natural Language Processing through a joint grant (EP/S022481/1) from UK Research and Innovation and the University of Edinburgh's School of Informatics and School of Philosophy, Psychology \& Language Sciences.

\bibliography{refs/main,refs/seeds,refs/lit,refs/all} 

\begin{thebibliography}{100}
\expandafter\ifx\csname url\endcsname\relax
  \def\url#1{\burl{#1}}\fi
\expandafter\ifx\csname urlprefix\endcsname\relax\def\urlprefix{URL }\fi
\providecommand{\bibinfo}[2]{#2}
\providecommand{\eprint}[2][]{\url{#2}}
\providecommand{\doi}[1]{\url{https://doi.org/#1}}
\bibcommenthead

\bibitem{kasirzadeh2024two}
\bibinfo{author}{Kasirzadeh, A.}
\newblock \bibinfo{title}{Two types of {AI} existential risk: Decisive and accumulative}.
\newblock \emph{\bibinfo{journal}{arXiv preprint arXiv:2401.07836}}  (\bibinfo{year}{2024}).

\bibitem{lazar2023ai}
\bibinfo{author}{Lazar, S.} \& \bibinfo{author}{Nelson, A.}
\newblock \bibinfo{title}{{AI} safety on whose terms?}
\newblock \emph{\bibinfo{journal}{Science}} \textbf{\bibinfo{volume}{381}}, \bibinfo{pages}{138--138} (\bibinfo{year}{2023}).

\bibitem{ahmed2023building}
\bibinfo{author}{Ahmed, S.}, \bibinfo{author}{Ja{\'z}wi{\'n}ska, K.}, \bibinfo{author}{Ahlawat, A.}, \bibinfo{author}{Winecoff, A.} \& \bibinfo{author}{Wang, M.}
\newblock \bibinfo{title}{Building the epistemic community of {AI} safety}.
\newblock \emph{\bibinfo{journal}{First Monday}}  (\bibinfo{year}{forthcoming}).
\newblock \urlprefix\url{https://papers.ssrn.com/sol3/papers.cfm?abstract_id=4641526}.

\bibitem{bostrom2002existential}
\bibinfo{author}{Bostrom, N.}
\newblock \bibinfo{title}{Existential risks: Analyzing human extinction scenarios and related hazards}.
\newblock \emph{\bibinfo{journal}{Journal of Evolution and Technology}} \textbf{\bibinfo{volume}{9}}, \bibinfo{pages}{1--30} (\bibinfo{year}{2002}).

\bibitem{bostrom2014superintelligence}
\bibinfo{author}{Bostrom, N.}
\newblock \emph{\bibinfo{title}{Superintelligence: Paths, dangers, strategies}}  (\bibinfo{publisher}{Oxford University Press}, \bibinfo{year}{2014}).

\bibitem{ord2020precipice}
\bibinfo{author}{Ord, T.}
\newblock \emph{\bibinfo{title}{The precipice: Existential risk and the future of humanity}}  (\bibinfo{publisher}{Hachette Books}, \bibinfo{year}{2020}).

\bibitem{Boston2015EffectiveAltruism}
\bibinfo{author}{{Boston Review}}.
\newblock \bibinfo{title}{The logic of effective altruism}.
\newblock \emph{\bibinfo{journal}{{Boston Review}}}  (\bibinfo{year}{2015}).
\newblock \urlprefix\url{https://www.bostonreview.net/forum/peter-singer-logic-effective-altruism/}.

\bibitem{gabriel2017effective}
\bibinfo{author}{Gabriel, I.}
\newblock \bibinfo{title}{Effective altruism and its critics}.
\newblock \emph{\bibinfo{journal}{Journal of Applied Philosophy}} \textbf{\bibinfo{volume}{34}}, \bibinfo{pages}{457--473} (\bibinfo{year}{2017}).

\bibitem{eisikovits2023ai}
\bibinfo{author}{Eisikovits, N.}
\newblock \bibinfo{title}{{AI} is an existential threat—just not the way you think}.
\newblock \bibinfo{howpublished}{\url{https://www.scientificamerican.com/article/ai-is-an-existential-threat-just-not-the-way-you-think/}} (\bibinfo{year}{2023}).
\newblock \bibinfo{note}{Accessed: July 12, 2023}.

\bibitem{CenterforAISafety2023}
\bibinfo{author}{{Center for AI Safety}}.
\newblock \bibinfo{title}{Statement on ai risk}.
\newblock \bibinfo{howpublished}{\url{https://www.safe.ai/work/statement-on-ai-risk}} (\bibinfo{year}{2023}).
\newblock \bibinfo{note}{Accessed on 2024-05-08}.

\bibitem{AISafetyTraining}
\bibinfo{author}{Training, A.~S.}
\newblock \bibinfo{title}{Ai safety training}.
\newblock \bibinfo{howpublished}{\url{https://aisafety.training/}} (\bibinfo{year}{2024}).
\newblock \bibinfo{note}{Accessed on 2024-05-08}.

\bibitem{wiki:aisafety}
\bibinfo{author}{{Wikipedia contributors}}.
\newblock \bibinfo{title}{{AI} safety --- wikipedia{,} the free encyclopedia} (\bibinfo{year}{2024}).
\newblock \urlprefix\url{https://en.wikipedia.org/wiki/AI_safety}.
\newblock \bibinfo{note}{[Online; accessed 21-January-2024]}.

\bibitem{aguera2023artificial}
\bibinfo{author}{Ag{\"u}era~y Arcas, B.} \& \bibinfo{author}{Norvig, P.}
\newblock \bibinfo{title}{Artificial general intelligence is already here}.
\newblock \emph{\bibinfo{journal}{Noema Magazine}}  (\bibinfo{year}{2023}).
\newblock \bibinfo{note}{Available at: \url{https://www.noemamag.com/artificial-general-intelligence-is-already-here/}}.

\bibitem{roose2023ai}
\bibinfo{author}{Roose, K.}
\newblock \bibinfo{title}{{A.I.} {Poses} ‘{Risk of Extinction}’, {Industry Leaders Warn}}.
\newblock \emph{\bibinfo{journal}{The New York Times}}  (\bibinfo{year}{2023}).
\newblock \urlprefix\url{https://www.nytimes.com/2023/05/30/technology/ai-threat-warning.html}.

\bibitem{bcs2023pmethics}
\bibinfo{author}{{BCS Comment}}.
\newblock \bibinfo{title}{{PM} should make ethics a priority at {AI} safety summit, say tech professionals} (\bibinfo{year}{2023}).
\newblock \urlprefix\url{https://www.bcs.org/articles-opinion-and-research/pm-should-make-ethics-a-priority-at-ai-safety-summit-say-tech-professionals/}.
\newblock \bibinfo{note}{Accessed: 27 January, 2024}.

\bibitem{gilardi2024we}
\bibinfo{author}{Gilardi, F.}, \bibinfo{author}{Kasirzadeh, A.}, \bibinfo{author}{Bernstein, A.}, \bibinfo{author}{Staab, S.} \& \bibinfo{author}{Gohdes, A.}
\newblock \bibinfo{title}{We need to understand the effect of narratives about generative ai}.
\newblock \emph{\bibinfo{journal}{Nature Human Behaviour}} \bibinfo{pages}{1--2} (\bibinfo{year}{2024}).

\bibitem{bender2023schism}
\bibinfo{author}{Bender, E.~M.}
\newblock \bibinfo{title}{Talking about a ‘schism’ is ahistorical} (\bibinfo{year}{2023}).
\newblock \urlprefix\url{https://medium.com/@emilymenonbender/talking-about-a-schism-is-ahistorical-3c454a77220f}.

\bibitem{krause2003aircraft}
\bibinfo{author}{Krause, S.~S.}
\newblock \emph{\bibinfo{title}{Aircraft safety}}  (\bibinfo{publisher}{McGraw-Hill Professional Publishing}, \bibinfo{year}{2003}).

\bibitem{boyd2017review}
\bibinfo{author}{Boyd, D.~D.}
\newblock \bibinfo{title}{A review of general aviation safety (1984--2017)}.
\newblock \emph{\bibinfo{journal}{Aerospace medicine and human performance}} \textbf{\bibinfo{volume}{88}}, \bibinfo{pages}{657--664} (\bibinfo{year}{2017}).

\bibitem{pifferi2003safety}
\bibinfo{author}{Pifferi, G.} \& \bibinfo{author}{Restani, P.}
\newblock \bibinfo{title}{The safety of pharmaceutical excipients}.
\newblock \emph{\bibinfo{journal}{Il Farmaco}} \textbf{\bibinfo{volume}{58}}, \bibinfo{pages}{541--550} (\bibinfo{year}{2003}).

\bibitem{leveson2012applying}
\bibinfo{author}{Leveson, N.} \emph{et~al.}
\newblock \bibinfo{title}{Applying system engineering to pharmaceutical safety}.
\newblock \emph{\bibinfo{journal}{Journal of Healthcare Engineering}} \textbf{\bibinfo{volume}{3}}, \bibinfo{pages}{391--414} (\bibinfo{year}{2012}).

\bibitem{de2019internet}
\bibinfo{author}{De~Kimpe, L.}, \bibinfo{author}{Walrave, M.}, \bibinfo{author}{Ponnet, K.} \& \bibinfo{author}{Van~Ouytsel, J.}
\newblock \bibinfo{title}{Internet safety}.
\newblock \emph{\bibinfo{journal}{The international encyclopedia of media literacy}} \bibinfo{pages}{1--11} (\bibinfo{year}{2019}).

\bibitem{salim2014cyber}
\bibinfo{author}{Salim, H.~M.}
\newblock \emph{\bibinfo{title}{Cyber safety: A systems thinking and systems theory approach to managing cyber security risks}}.
\newblock Ph.D. thesis, \bibinfo{school}{Massachusetts Institute of Technology} (\bibinfo{year}{2014}).

\bibitem{leveson2016engineering}
\bibinfo{author}{Leveson, N.~G.}
\newblock \emph{\bibinfo{title}{Engineering a safer world: Systems thinking applied to safety}}  (\bibinfo{publisher}{The MIT Press}, \bibinfo{year}{2016}).

\bibitem{varshney2016engineering}
\bibinfo{author}{Varshney, K.~R.}
\newblock \emph{\bibinfo{title}{Engineering safety in machine learning}}, \bibinfo{pages}{1--5} (\bibinfo{organization}{IEEE}, \bibinfo{year}{2016}).

\bibitem{rismani2023plane}
\bibinfo{author}{Rismani, S.} \emph{et~al.}
\newblock \emph{\bibinfo{title}{From plane crashes to algorithmic harm: applicability of safety engineering frameworks for responsible ml}}, \bibinfo{pages}{1--18} (\bibinfo{year}{2023}).

\bibitem{dobbe2022system}
\bibinfo{author}{Dobbe, R.}
\newblock \emph{\bibinfo{title}{System safety and artificial intelligence}}, \bibinfo{pages}{1584--1584} (\bibinfo{year}{2022}).

\bibitem{rismani2023beyond}
\bibinfo{author}{Rismani, S.} \emph{et~al.}
\newblock \emph{\bibinfo{title}{Beyond the ml model: Applying safety engineering frameworks to text-to-image development}}, \bibinfo{pages}{70--83} (\bibinfo{year}{2023}).

\bibitem{amodei2016concrete}
\bibinfo{author}{Amodei, D.} \emph{et~al.}
\newblock \bibinfo{title}{Concrete problems in {AI} safety}.
\newblock \emph{\bibinfo{journal}{arXiv preprint arXiv:1606.06565}}  (\bibinfo{year}{2016}).

\bibitem{raji2023concrete}
\bibinfo{author}{Raji, I.~D.} \& \bibinfo{author}{Dobbe, R.}
\newblock \bibinfo{title}{Concrete problems in {AI} safety, revisited}.
\newblock \emph{\bibinfo{journal}{arXiv preprint arXiv:2401.10899}}  (\bibinfo{year}{2023}).

\bibitem{kitchenhamGuidelinesPerformingSystematic2007}
\bibinfo{author}{Kitchenham, B.} \& \bibinfo{author}{Charters, S.}
\newblock \bibinfo{title}{Guidelines for performing {{Systematic Literature Reviews}} in {{Software Engineering}}}.
\newblock \bibinfo{type}{{{EBSE Technical Report}}} \bibinfo{number}{EBSE-2007-01}, \bibinfo{institution}{{School of Computer Science and Mathematics, Keele University}}, \bibinfo{address}{{Keele, UK}} (\bibinfo{year}{2007}).

\bibitem{wohlinGuidelinesSnowballingSystematic2014}
\bibinfo{author}{Wohlin, C.}
\newblock \emph{\bibinfo{title}{Guidelines for snowballing in systematic literature studies and a replication in software engineering}}, {{EASE}} '14, \bibinfo{pages}{1--10} (\bibinfo{publisher}{{Association for Computing Machinery}}, \bibinfo{address}{{New York, NY, USA}}, \bibinfo{year}{2014}).

\bibitem{irvingAISafetyDebate2018}
\bibinfo{author}{Irving, G.}, \bibinfo{author}{Christiano, P.} \& \bibinfo{author}{Amodei, D.}
\newblock \bibinfo{title}{{{AI}} safety via debate}.
\newblock \emph{\bibinfo{journal}{{arXiv}}}  (\bibinfo{year}{2018}).

\bibitem{ngAlgorithmsInverseReinforcement2000}
\bibinfo{author}{Ng, A.~Y.} \& \bibinfo{author}{Russell, S.~J.}
\newblock \emph{\bibinfo{title}{Algorithms for {{Inverse Reinforcement Learning}}}}, {{ICML}} '00, \bibinfo{pages}{663--670} (\bibinfo{publisher}{{Morgan Kaufmann Publishers Inc.}}, \bibinfo{address}{{San Francisco, CA, USA}}, \bibinfo{year}{2000}).

\bibitem{hendrycksAligningAIShared2021}
\bibinfo{author}{Hendrycks, D.} \emph{et~al.}
\newblock \emph{\bibinfo{title}{Aligning {{AI With Shared Human Values}}}} (\bibinfo{address}{{ICLR}}, \bibinfo{year}{2021}).
\newblock \eprint{2008.02275}.

\bibitem{yampolskiyArtificialIntelligenceSafety2016}
\bibinfo{author}{Yampolskiy, R.~V.}
\newblock \bibinfo{title}{Artificial {{Intelligence Safety}} and {{Cybersecurity}}: A {{Timeline}} of {{AI Failures}}}.
\newblock \emph{\bibinfo{journal}{{arXiv}}}  (\bibinfo{year}{2016}).

\bibitem{hadfield-menellCooperativeInverseReinforcement2016}
\bibinfo{author}{{Hadfield-Menell}, D.}, \bibinfo{author}{Russell, S.~J.}, \bibinfo{author}{Abbeel, P.} \& \bibinfo{author}{Dragan, A.}
\newblock \emph{\bibinfo{title}{Cooperative {{Inverse Reinforcement Learning}}}}, Vol.~\bibinfo{volume}{29} (\bibinfo{publisher}{{Curran Associates, Inc.}}, \bibinfo{year}{2016}).

\bibitem{xuMachineUnlearningSurvey2023}
\bibinfo{author}{Xu, H.}, \bibinfo{author}{Zhu, T.}, \bibinfo{author}{Zhang, L.}, \bibinfo{author}{Zhou, W.} \& \bibinfo{author}{Yu, P.~S.}
\newblock \bibinfo{title}{Machine {{Unlearning}}: {{A Survey}}}.
\newblock \emph{\bibinfo{journal}{ACM Computing Surveys}} \textbf{\bibinfo{volume}{56}}, \bibinfo{pages}{9:1--9:36} (\bibinfo{year}{2023}).

\bibitem{russellResearchPrioritiesRobust2016}
\bibinfo{author}{Russell, S.}, \bibinfo{author}{Dewey, D.} \& \bibinfo{author}{Tegmark, M.}
\newblock \bibinfo{title}{Research {{Priorities}} for {{Robust}} and {{Beneficial Artificial Intelligence}}}.
\newblock \emph{\bibinfo{journal}{AI magazine}} \textbf{\bibinfo{volume}{36}}, \bibinfo{pages}{105--114} (\bibinfo{year}{2015}).

\bibitem{willersSafetyConcernsMitigation2020}
\bibinfo{author}{Willers, O.}, \bibinfo{author}{Sudholt, S.}, \bibinfo{author}{Raafatnia, S.} \& \bibinfo{author}{Abrecht, S.}
\newblock \bibinfo{editor}{Casimiro, A.}, \bibinfo{editor}{Ortmeier, F.}, \bibinfo{editor}{Schoitsch, E.}, \bibinfo{editor}{Bitsch, F.} \& \bibinfo{editor}{Ferreira, P.} (eds) \emph{\bibinfo{title}{Safety {{Concerns}} and {{Mitigation Approaches Regarding}} the {{Use}} of {{Deep Learning}} in {{Safety-Critical Perception Tasks}}}}.
\newblock (eds \bibinfo{editor}{Casimiro, A.}, \bibinfo{editor}{Ortmeier, F.}, \bibinfo{editor}{Schoitsch, E.}, \bibinfo{editor}{Bitsch, F.} \& \bibinfo{editor}{Ferreira, P.}) \emph{\bibinfo{booktitle}{Computer {{Safety}}, {{Reliability}}, and {{Security}}. {{SAFECOMP}} 2020 {{Workshops}}}}, Lecture {{Notes}} in {{Computer Science}}, \bibinfo{pages}{336--350} (\bibinfo{publisher}{{Springer International Publishing}}, \bibinfo{address}{{Cham}}, \bibinfo{year}{2020}).

\bibitem{mohseniTaxonomyMachineLearning2022}
\bibinfo{author}{Mohseni, S.} \emph{et~al.}
\newblock \bibinfo{title}{Taxonomy of {{Machine Learning Safety}}: {{A Survey}} and {{Primer}}}.
\newblock \emph{\bibinfo{journal}{ACM Computing Surveys}} \textbf{\bibinfo{volume}{55}}, \bibinfo{pages}{1--38} (\bibinfo{year}{2022}).

\bibitem{hendrycksUnsolvedProblemsML2022}
\bibinfo{author}{Hendrycks, D.}, \bibinfo{author}{Carlini, N.}, \bibinfo{author}{Schulman, J.} \& \bibinfo{author}{Steinhardt, J.}
\newblock \bibinfo{title}{Unsolved {{Problems}} in {{ML Safety}}}.
\newblock \emph{\bibinfo{journal}{{arXiv}}}  (\bibinfo{year}{2022}).

\bibitem{boyatzis1998transforming}
\bibinfo{author}{Boyatzis, R.~E.}
\newblock \emph{\bibinfo{title}{Transforming qualitative information: Thematic analysis and code development}}  (\bibinfo{publisher}{sage}, \bibinfo{year}{1998}).

\bibitem{vaneckSoftwareSurveyVOSviewer2010}
\bibinfo{author}{{van Eck}, N.~J.} \& \bibinfo{author}{Waltman, L.}
\newblock \bibinfo{title}{Software survey: {{VOSviewer}}, a computer program for bibliometric mapping}.
\newblock \emph{\bibinfo{journal}{Scientometrics}} \textbf{\bibinfo{volume}{84}}, \bibinfo{pages}{523--538} (\bibinfo{year}{2010}).

\bibitem{oster2013analyzing}
\bibinfo{author}{Oster~Jr, C.~V.}, \bibinfo{author}{Strong, J.~S.} \& \bibinfo{author}{Zorn, C.~K.}
\newblock \bibinfo{title}{Analyzing aviation safety: Problems, challenges, opportunities}.
\newblock \emph{\bibinfo{journal}{Research in transportation economics}} \textbf{\bibinfo{volume}{43}}, \bibinfo{pages}{148--164} (\bibinfo{year}{2013}).

\bibitem{donaldson2000err}
\bibinfo{author}{Donaldson, M.~S.}, \bibinfo{author}{Corrigan, J.~M.} \& \bibinfo{author}{Kohn, L.~T.}
\newblock \bibinfo{title}{To err is human: building a safer health system}  (\bibinfo{year}{2000}).

\bibitem{bates2023safety}
\bibinfo{author}{Bates, D.~W.} \emph{et~al.}
\newblock \bibinfo{title}{The safety of inpatient health care}.
\newblock \emph{\bibinfo{journal}{New England Journal of Medicine}} \textbf{\bibinfo{volume}{388}}, \bibinfo{pages}{142--153} (\bibinfo{year}{2023}).

\bibitem{marais2004beyond}
\bibinfo{author}{Marais, K.}, \bibinfo{author}{Dulac, N.}, \bibinfo{author}{Leveson, N.} \emph{et~al.}
\newblock \emph{\bibinfo{title}{Beyond normal accidents and high reliability organizations: The need for an alternative approach to safety in complex systems}}, \bibinfo{pages}{1--16} (\bibinfo{organization}{Citeseer}, \bibinfo{year}{2004}).

\bibitem{griffor2016handbook}
\bibinfo{author}{Griffor, E.}
\newblock \emph{\bibinfo{title}{Handbook of system safety and security: cyber risk and risk management, cyber security, threat analysis, functional safety, software systems, and cyber physical systems}}  (\bibinfo{publisher}{Syngress}, \bibinfo{year}{2016}).

\bibitem{prasad2020cyber}
\bibinfo{author}{Prasad, R.} \& \bibinfo{author}{Rohokale, V.}
\newblock \emph{\bibinfo{title}{Cyber security: the lifeline of information and communication technology}}  (\bibinfo{publisher}{Springer}, \bibinfo{year}{2020}).

\bibitem{soaresCorrigibility2015}
\bibinfo{author}{Soares, N.}, \bibinfo{author}{Fallenstein, B.}, \bibinfo{author}{Yudkowsky, E.} \& \bibinfo{author}{Armstrong, S.}
\newblock \emph{\bibinfo{title}{Corrigibility}} (\bibinfo{address}{{Austin, Texas, USA}}, \bibinfo{year}{2015}).

\bibitem{ringDelusionSurvivalIntelligent2011}
\bibinfo{author}{Ring, M.} \& \bibinfo{author}{Orseau, L.}
\newblock \bibinfo{editor}{Schmidhuber, J.}, \bibinfo{editor}{Th{\'o}risson, K.~R.} \& \bibinfo{editor}{Looks, M.} (eds) \emph{\bibinfo{title}{Delusion, {{Survival}}, and {{Intelligent Agents}}}}.
\newblock (eds \bibinfo{editor}{Schmidhuber, J.}, \bibinfo{editor}{Th{\'o}risson, K.~R.} \& \bibinfo{editor}{Looks, M.}) \emph{\bibinfo{booktitle}{Artificial {{General Intelligence}}}}, Lecture {{Notes}} in {{Computer Science}}, \bibinfo{pages}{11--20} (\bibinfo{publisher}{{Springer}}, \bibinfo{address}{{Berlin, Heidelberg}}, \bibinfo{year}{2011}).

\bibitem{ganinDomainAdversarialTrainingNeural2017}
\bibinfo{author}{Ganin, Y.} \emph{et~al.}
\newblock \bibinfo{title}{ in \textit{Domain-{{Adversarial Training}} of {{Neural Networks}}}} (ed.\bibinfo{editor}{Csurka, G.}) \emph{\bibinfo{booktitle}{Domain {{Adaptation}} in {{Computer Vision Applications}}}} \bibinfo{pages}{189--209} (\bibinfo{publisher}{{Springer International Publishing}}, \bibinfo{address}{{Cham}}, \bibinfo{year}{2017}).

\bibitem{balajiMetaRegDomainGeneralization2018}
\bibinfo{author}{Balaji, Y.}, \bibinfo{author}{Sankaranarayanan, S.} \& \bibinfo{author}{Chellappa, R.}
\newblock \emph{\bibinfo{title}{{{MetaReg}}: {{Towards Domain Generalization}} using {{Meta-Regularization}}}}, Vol.~\bibinfo{volume}{31} (\bibinfo{publisher}{{Curran Associates, Inc.}}, \bibinfo{year}{2018}).

\bibitem{goodfellowExplainingHarnessingAdversarial2015}
\bibinfo{author}{Goodfellow, I.~J.}, \bibinfo{author}{Shlens, J.} \& \bibinfo{author}{Szegedy, C.}
\newblock \emph{\bibinfo{title}{Explaining and {{Harnessing Adversarial Examples}}}} (\bibinfo{year}{2015}).
\newblock \eprint{1412.6572}.

\bibitem{madryDeepLearningModels2019}
\bibinfo{author}{Madry, A.}, \bibinfo{author}{Makelov, A.}, \bibinfo{author}{Schmidt, L.}, \bibinfo{author}{Tsipras, D.} \& \bibinfo{author}{Vladu, A.}
\newblock \bibinfo{title}{Towards {{Deep Learning Models Resistant}} to {{Adversarial Attacks}}}.
\newblock \emph{\bibinfo{journal}{{arXiv}}}  (\bibinfo{year}{2019}).

\bibitem{rayBenchmarkingSafeExploration2023}
\bibinfo{author}{Ray, A.}, \bibinfo{author}{Achiam, J.} \& \bibinfo{author}{Amodei, D.}
\newblock \bibinfo{title}{Benchmarking {{Safe Exploration}} in {{Deep Reinforcement Learning}}}.
\newblock \emph{\bibinfo{journal}{{OpenAI}}}  (\bibinfo{year}{2023}).
\newblock \bibinfo{note}{Unpublished article}.

\bibitem{turchettaSafeExplorationInteractive2019}
\bibinfo{author}{Turchetta, M.}, \bibinfo{author}{Berkenkamp, F.} \& \bibinfo{author}{Krause, A.}
\newblock \bibinfo{editor}{Wallach, H.} \emph{et~al.} (eds) \emph{\bibinfo{title}{Safe {{Exploration}} for {{Interactive Machine Learning}}}}.
\newblock (eds \bibinfo{editor}{Wallach, H.} \emph{et~al.}) \emph{\bibinfo{booktitle}{{{Advances in Neural Information Processing Systems}} 32 ({{NIPS}} 2019)}}, Vol.~\bibinfo{volume}{32} (\bibinfo{year}{2019}).

\bibitem{ouyangTrainingLanguageModels2022}
\bibinfo{author}{Ouyang, L.} \emph{et~al.}
\newblock \bibinfo{title}{Training language models to follow instructions with human feedback}.
\newblock \emph{\bibinfo{journal}{{arXiv}}}  (\bibinfo{year}{2022}).
\newblock \bibinfo{note}{{OpenAI}}.

\bibitem{abbeelApprenticeshipLearningInverse2004}
\bibinfo{author}{Abbeel, P.} \& \bibinfo{author}{Ng, A.~Y.}
\newblock \emph{\bibinfo{title}{Apprenticeship learning via inverse reinforcement learning}}, {{ICML}} '04, \bibinfo{pages}{1} (\bibinfo{publisher}{{Association for Computing Machinery}}, \bibinfo{address}{{New York, NY, USA}}, \bibinfo{year}{2004}).

\bibitem{izzoApproximateDataDeletion2021}
\bibinfo{author}{Izzo, Z.}, \bibinfo{author}{Smart, M.~A.}, \bibinfo{author}{Chaudhuri, K.} \& \bibinfo{author}{Zou, J.}
\newblock \emph{\bibinfo{title}{Approximate {{Data Deletion}} from {{Machine Learning Models}}}}, \bibinfo{pages}{2008--2016} (\bibinfo{publisher}{{PMLR}}, \bibinfo{year}{2021}).

\bibitem{wuDeltaGradRapidRetraining2020}
\bibinfo{author}{Wu, Y.}, \bibinfo{author}{Dobriban, E.} \& \bibinfo{author}{Davidson, S.}
\newblock \emph{\bibinfo{title}{{{DeltaGrad}}: {{Rapid}} retraining of machine learning models}}, \bibinfo{pages}{10355--10366} (\bibinfo{publisher}{{PMLR}}, \bibinfo{year}{2020}).

\bibitem{adebayoSanityChecksSaliency2018}
\bibinfo{author}{Adebayo, J.} \emph{et~al.}
\newblock \emph{\bibinfo{title}{Sanity {{Checks}} for {{Saliency Maps}}}}, Vol.~\bibinfo{volume}{31} (\bibinfo{publisher}{{Curran Associates, Inc.}}, \bibinfo{year}{2018}).

\bibitem{kimInterpretabilityFeatureAttribution2018}
\bibinfo{author}{Kim, B.} \emph{et~al.}
\newblock \emph{\bibinfo{title}{Interpretability beyond feature attribution: Quantitative testing with concept activation vectors ({TCAV})}}, \bibinfo{pages}{2668--2677} (\bibinfo{organization}{PMLR}, \bibinfo{year}{2018}).

\bibitem{hendrycksBaselineDetectingMisclassified2018}
\bibinfo{author}{Hendrycks, D.} \& \bibinfo{author}{Gimpel, K.}
\newblock \emph{\bibinfo{title}{A {{Baseline}} for {{Detecting Misclassified}} and {{Out-of-Distribution Examples}} in {{Neural Networks}}}} (\bibinfo{publisher}{{arXiv}}, \bibinfo{year}{2018}).
\newblock \eprint{1610.02136}.

\bibitem{hendrycksBenchmarkingNeuralNetwork2019}
\bibinfo{author}{Hendrycks, D.} \& \bibinfo{author}{Dietterich, T.}
\newblock \emph{\bibinfo{title}{Benchmarking {{Neural Network Robustness}} to {{Common Corruptions}} and {{Perturbations}}}} (\bibinfo{publisher}{{arXiv}}, \bibinfo{year}{2019}).
\newblock \eprint{1903.12261}.

\bibitem{ortega2018building}
\bibinfo{author}{Ortega, P.~A.}, \bibinfo{author}{Maini, V.} \& \bibinfo{author}{the DeepMind~safety team}.
\newblock \bibinfo{title}{Building safe artificial intelligence: specification, robustness, and assurance}.
\newblock \bibinfo{howpublished}{\url{https://deepmindsafetyresearch.medium.com/building-safe-artificial-intelligence-52f5f75058f1}} (\bibinfo{year}{2018}).
\newblock \bibinfo{note}{Accessed: 2024-05-13}.

\bibitem{mengDistantlySupervisedNamedEntity2021}
\bibinfo{author}{Meng, Y.} \emph{et~al.}
\newblock \emph{\bibinfo{title}{Distantly-{{Supervised Named Entity Recognition}} with {{Noise-Robust Learning}} and {{Language Model Augmented Self-Training}}}}, \bibinfo{pages}{10367--10378} (\bibinfo{year}{2021}).

\bibitem{wangRobustAutomatedMachine2021}
\bibinfo{author}{Wang, K.} \& \bibinfo{author}{Guo, P.}
\newblock \bibinfo{title}{A {{Robust Automated Machine Learning System}} with {{Pseudoinverse Learning}}}.
\newblock \emph{\bibinfo{journal}{Cognitive Computation}} \textbf{\bibinfo{volume}{13}}, \bibinfo{pages}{724--735} (\bibinfo{year}{2021}).

\bibitem{cappozzoRobustApproachModelbased2020}
\bibinfo{author}{Cappozzo, A.}, \bibinfo{author}{Greselin, F.} \& \bibinfo{author}{Murphy, T.~B.}
\newblock \bibinfo{title}{A robust approach to model-based classification based on trimming and constraints: Semi-supervised learning in presence of outliers and label noise}.
\newblock \emph{\bibinfo{journal}{Advances in Data Analysis and Classification}} \textbf{\bibinfo{volume}{14}}, \bibinfo{pages}{327--354} (\bibinfo{year}{2020}).

\bibitem{liRobustSupervisedSubspace2021}
\bibinfo{author}{Li, W.} \& \bibinfo{author}{Wang, Y.}
\newblock \bibinfo{title}{A robust supervised subspace learning approach for output-relevant prediction and detection against outliers}.
\newblock \emph{\bibinfo{journal}{Journal of Process Control}} \textbf{\bibinfo{volume}{106}}, \bibinfo{pages}{184--194} (\bibinfo{year}{2021}).

\bibitem{curiCombiningPessimismOptimism2021}
\bibinfo{author}{Curi, S.}, \bibinfo{author}{Bogunovic, I.} \& \bibinfo{author}{Krause, A.}
\newblock \emph{\bibinfo{title}{Combining {{Pessimism}} with {{Optimism}} for {{Robust}} and {{Efficient Model-Based Deep Reinforcement Learning}}}}, Vol. \bibinfo{volume}{139}, \bibinfo{pages}{2254--2264} (\bibinfo{year}{2021}).

\bibitem{dobbeHardChoicesArtificial2021}
\bibinfo{author}{Dobbe, R.}, \bibinfo{author}{Gilbert, {\relax TK}.} \& \bibinfo{author}{Mintz, Y.}
\newblock \bibinfo{title}{Hard choices in artificial intelligence}.
\newblock \emph{\bibinfo{journal}{Artificial Intelligence}} \textbf{\bibinfo{volume}{300}} (\bibinfo{year}{2021}).

\bibitem{dworkPrivacypreservingPrediction2018}
\bibinfo{author}{Dwork, C.} \& \bibinfo{author}{Feldman, V.}
\newblock \emph{\bibinfo{title}{Privacy-preserving {{Prediction}}}}, \bibinfo{pages}{1693--1702} (\bibinfo{publisher}{{PMLR}}, \bibinfo{year}{2018}).

\bibitem{elhageMathematicalFrameworkTransformer2021}
\bibinfo{author}{Elhage, N.} \emph{et~al.}
\newblock \bibinfo{title}{A mathematical framework for transformer circuits}.
\newblock \emph{\bibinfo{journal}{Transformer Circuits Thread}}  (\bibinfo{year}{2021}).
\newblock \bibinfo{note}{Unpublished article}.

\bibitem{kimDisentanglingFactorising2019}
\bibinfo{author}{Kim, H.} \& \bibinfo{author}{Mnih, A.}
\newblock \bibinfo{title}{Disentangling by {{Factorising}}}.
\newblock \emph{\bibinfo{journal}{{arXiv}}}  (\bibinfo{year}{2019}).

\bibitem{wardAssuranceCasePattern2020}
\bibinfo{author}{Ward, F.} \& \bibinfo{author}{Habli, I.}
\newblock \emph{\bibinfo{title}{An {{Assurance Case Pattern}} for the {{Interpretability}} of {{Machine Learning}} in {{Safety-Critical Systems}}}}, Vol. \bibinfo{volume}{12235 LNCS}, \bibinfo{pages}{395--407} (\bibinfo{year}{2020}).

\bibitem{gyevnar2023transparencyGap}
\bibinfo{author}{Gyevnar, B.}, \bibinfo{author}{Ferguson, N.} \& \bibinfo{author}{Schafer, B.}
\newblock \emph{\bibinfo{title}{Bridging the transparency gap: What can explainable ai learn from the ai act?}}, \bibinfo{pages}{964--971} (\bibinfo{organization}{IOS Press}, \bibinfo{year}{2023}).

\bibitem{reimannSafeDSDomainSpecific2023}
\bibinfo{author}{Reimann, L.} \& \bibinfo{author}{{Kniesel-W{\"u}nsche}, G.}
\newblock \emph{\bibinfo{title}{Safe-{{DS}}: {{A Domain Specific Language}} to {{Make Data Science Safe}}}}, \bibinfo{pages}{72--77} (\bibinfo{year}{2023}).

\bibitem{deyMultilayeredCollaborativeFramework2023}
\bibinfo{author}{Dey, S.} \& \bibinfo{author}{Lee, S.-W.}
\newblock \emph{\bibinfo{title}{A {{Multi-layered Collaborative Framework}} for {{Evidence-driven Data Requirements Engineering}} for {{Machine Learning-based Safety-critical Systems}}}}, \bibinfo{pages}{1404--1413} (\bibinfo{year}{2023}).

\bibitem{weiModelSelectionApproach2022}
\bibinfo{author}{Wei, C.-Y.}, \bibinfo{author}{Dann, C.} \& \bibinfo{author}{Zimmert, J.}
\newblock \emph{\bibinfo{title}{A {{Model Selection Approach}} for {{Corruption Robust Reinforcement Learning}}}}, Vol. \bibinfo{volume}{167}, \bibinfo{pages}{1043--1096} (\bibinfo{year}{2022}).

\bibitem{ghoshDeploymentRobustCooperative2020a}
\bibinfo{author}{Ghosh, A.}, \bibinfo{author}{Tschiatschek, S.}, \bibinfo{author}{Mahdavi, H.} \& \bibinfo{author}{Singla, A.}
\newblock \bibinfo{title}{Towards {{Deployment}} of {{Robust Cooperative AI Agents}}: {{An Algorithmic Framework}} for {{Learning Adaptive Policies}}}.
\newblock \emph{\bibinfo{journal}{New Zealand}}  (\bibinfo{year}{2020}).

\bibitem{everittAvoidingWireheadingValue2016}
\bibinfo{author}{Everitt, T.} \& \bibinfo{author}{Hutter, M.}
\newblock \bibinfo{title}{Avoiding {{Wireheading}} with {{Value Reinforcement Learning}}} (\bibinfo{year}{2016}).
\newblock \eprint{1605.03143}.

\bibitem{mesaoptimisation2019}
\bibinfo{author}{Hubinger, E.}, \bibinfo{author}{van Merwijk, C.}, \bibinfo{author}{Mikulik, V.}, \bibinfo{author}{Skalse, J.} \& \bibinfo{author}{Garrabrant, S.}
\newblock \bibinfo{title}{Risks from learned optimization in advanced machine learning systems}.
\newblock \emph{\bibinfo{journal}{CoRR}} \textbf{\bibinfo{volume}{abs/1906.01820}} (\bibinfo{year}{2019}).
\newblock \urlprefix\url{http://arxiv.org/abs/1906.01820}.

\bibitem{pistonoUnethicalResearchHow2016}
\bibinfo{author}{Pistono, F.} \& \bibinfo{author}{Yampolskiy, R.~V.}
\newblock \emph{\bibinfo{title}{The Age of Artificial Intelligence}}, Ch. \bibinfo{chapter}{Unethical {{Research}}: {{How}} to {{Create}} a {{Malevolent Artificial Intelligence}}} (\bibinfo{publisher}{Vernon Press Wilmington, DE, USA}, \bibinfo{year}{2016}).

\bibitem{picardiAssuranceArgumentPatterns2020}
\bibinfo{author}{Picardi, C.}, \bibinfo{author}{Paterson, C.}, \bibinfo{author}{Hawkins, R.}, \bibinfo{author}{Calinescu, R.} \& \bibinfo{author}{Habli, I.}
\newblock \emph{\bibinfo{title}{Assurance argument patterns and processes for machine learning in safety-related systems}}, Vol. \bibinfo{volume}{2560}, \bibinfo{pages}{23--30} (\bibinfo{year}{2020}).

\bibitem{wabersichProbabilisticModelPredictive2022}
\bibinfo{author}{Wabersich, {\relax KJ}.}, \bibinfo{author}{Hewing, L.}, \bibinfo{author}{Carron, A.} \& \bibinfo{author}{Zeilinger, {\relax MN}.}
\newblock \bibinfo{title}{Probabilistic {{Model Predictive Safety Certification}} for {{Learning-Based Control}}}.
\newblock \emph{\bibinfo{journal}{IEEE Transactions on Automatic Control}} \textbf{\bibinfo{volume}{67}}, \bibinfo{pages}{176--188} (\bibinfo{year}{2022}).

\bibitem{wenConstrainedCrossEntropyMethod2021}
\bibinfo{author}{Wen, M.} \& \bibinfo{author}{Topcu, U.}
\newblock \bibinfo{title}{Constrained {{Cross-Entropy Method}} for {{Safe Reinforcement Learning}}}.
\newblock \emph{\bibinfo{journal}{IEEE Transactions on Automatic Control}} \textbf{\bibinfo{volume}{66}}, \bibinfo{pages}{3123--3137} (\bibinfo{year}{2021}).

\bibitem{zanella-beguelinAnalyzingInformationLeakage2020}
\bibinfo{author}{{Zanella-B{\'e}guelin}, S.} \emph{et~al.}
\newblock \emph{\bibinfo{title}{Analyzing {{Information Leakage}} of {{Updates}} to {{Natural Language Models}}}}, \bibinfo{pages}{363--375} (\bibinfo{year}{2020}).
\newblock \eprint{1912.07942}.

\bibitem{wangDirichletProcessMixture2022}
\bibinfo{author}{Wang, Z.}, \bibinfo{author}{Chen, C.} \& \bibinfo{author}{Dong, D.}
\newblock \bibinfo{title}{A {{Dirichlet Process Mixture}} of {{Robust Task Models}} for {{Scalable Lifelong Reinforcement Learning}}}.
\newblock \emph{\bibinfo{journal}{IEEE Transactions on Cybernetics}} \bibinfo{pages}{1--12} (\bibinfo{year}{2022}).

\bibitem{zouUniversalTransferableAdversarial2023}
\bibinfo{author}{Zou, A.} \emph{et~al.}
\newblock \bibinfo{title}{Universal and {{Transferable Adversarial Attacks}} on {{Aligned Language Models}}}.
\newblock \emph{\bibinfo{journal}{{arXiv}}}  (\bibinfo{year}{2023}).

\bibitem{ilyasAdversarialExamplesAre2019a}
\bibinfo{author}{Ilyas, A.} \emph{et~al.}
\newblock \emph{\bibinfo{title}{Adversarial {{Examples Are Not Bugs}}, {{They Are Features}}}}, Vol.~\bibinfo{volume}{32} (\bibinfo{publisher}{{Curran Associates, Inc.}}, \bibinfo{year}{2019}).

\bibitem{heNotAllParameters2022}
\bibinfo{author}{He, {\relax RD}.}, \bibinfo{author}{Han, {\relax ZY}.}, \bibinfo{author}{Yang, Y.} \& \bibinfo{author}{Yin, {\relax YL}.}
\newblock \emph{\bibinfo{title}{Not {{All Parameters Should Be Treated Equally}}: {{Deep Safe Semi-supervised Learning}} under {{Class Distribution Mismatch}}}}, \bibinfo{pages}{6874--6883} (\bibinfo{year}{2022}).

\bibitem{aghakhaniBullseyePolytopeScalable2021}
\bibinfo{author}{Aghakhani, H.}, \bibinfo{author}{Meng, D.}, \bibinfo{author}{Wang, Y.-X.}, \bibinfo{author}{Kruegel, C.} \& \bibinfo{author}{Vigna, G.}
\newblock \emph{\bibinfo{title}{Bullseye polytope: {{A}} scalable clean-label poisoning attack with improved transferability}}, \bibinfo{pages}{159--178} (\bibinfo{year}{2021}).

\bibitem{liuBackdoorDefenseMachine2022}
\bibinfo{author}{Liu, Y.} \emph{et~al.}
\newblock \emph{\bibinfo{title}{Backdoor {{Defense}} with {{Machine Unlearning}}}}, \bibinfo{pages}{280--289} (\bibinfo{publisher}{{IEEE Press}}, \bibinfo{address}{{London, United Kingdom}}, \bibinfo{year}{2022}).

\bibitem{meinkeNeuralNetworksThat2020}
\bibinfo{author}{Meinke, A.} \& \bibinfo{author}{Hein, M.}
\newblock \bibinfo{title}{Towards neural networks that provably know when they don't know} (\bibinfo{year}{2020}).
\newblock \eprint{1909.12180}.

\bibitem{abdelfattahRobustPolicyBootstrapping2020}
\bibinfo{author}{Abdelfattah, S.}, \bibinfo{author}{Kasmarik, K.} \& \bibinfo{author}{Hu, J.}
\newblock \bibinfo{title}{A robust policy bootstrapping algorithm for multi-objective reinforcement learning in non-stationary environments}.
\newblock \emph{\bibinfo{journal}{Adaptive Behavior}} \textbf{\bibinfo{volume}{28}}, \bibinfo{pages}{273--292} (\bibinfo{year}{2020}).

\bibitem{djeumouTaskGuidedInverseReinforcement2021}
\bibinfo{author}{Djeumou, F.}, \bibinfo{author}{Cubuktepe, M.}, \bibinfo{author}{Lennon, C.} \& \bibinfo{author}{Topcu, U.}
\newblock \bibinfo{title}{Task-{{Guided Inverse Reinforcement Learning Under Partial Information}}} (\bibinfo{year}{2021}).
\newblock \eprint{2105.14073}.

\bibitem{yampolskiyLeakproofingSingularityArtificial2012}
\bibinfo{author}{Yampolskiy, R.~V.}
\newblock \bibinfo{title}{Leakproofing the {{Singularity}}: {{Artificial}} intelligence confinement problem}.
\newblock \emph{\bibinfo{journal}{Journal of Consciousness Studies}} \textbf{\bibinfo{volume}{19}}, \bibinfo{pages}{194--214} (\bibinfo{year}{2012}).

\bibitem{yampolskiyArtificialIntelligenceSafety2013}
\bibinfo{author}{Yampolskiy, R.~V.}
\newblock \bibinfo{title}{ in \textit{Artificial {{Intelligence Safety Engineering}}: {{Why Machine Ethics Is}} a {{Wrong Approach}}}} (ed.\bibinfo{editor}{M{\"u}ller, V.~C.}) \emph{\bibinfo{booktitle}{Philosophy and {{Theory}} of {{Artificial Intelligence}}}} Studies in {{Applied Philosophy}}, {{Epistemology}} and {{Rational Ethics}}, \bibinfo{pages}{389--396} (\bibinfo{publisher}{{Springer}}, \bibinfo{address}{{Berlin, Heidelberg}}, \bibinfo{year}{2013}).

\bibitem{yampolskiyTaxonomyPathwaysDangerous2015}
\bibinfo{author}{Yampolskiy, R.~V.}
\newblock \bibinfo{title}{Taxonomy of {{Pathways}} to {{Dangerous AI}}} (\bibinfo{year}{2015}).
\newblock \eprint{1511.03246}.

\bibitem{wheatley2016reassessing}
\bibinfo{author}{Wheatley, S.}, \bibinfo{author}{Sovacool, B.~K.} \& \bibinfo{author}{Sornette, D.}
\newblock \bibinfo{title}{Reassessing the safety of nuclear power}.
\newblock \emph{\bibinfo{journal}{Energy Research \& Social Science}} \textbf{\bibinfo{volume}{15}}, \bibinfo{pages}{96--100} (\bibinfo{year}{2016}).

\bibitem{tayStudyRealtimeArtificial1998}
\bibinfo{author}{Tay, {\relax EB}.}, \bibinfo{author}{Gan, {\relax OP}.} \& \bibinfo{author}{Ho, {\relax WK}.}
\newblock \bibinfo{editor}{Rauch, {\relax HE}.} (ed.) \emph{\bibinfo{title}{A study on real-time artificial intelligence}}.
\newblock (ed.\bibinfo{editor}{Rauch, {\relax HE}.}) \emph{\bibinfo{booktitle}{Artificial Intelligence in Real-Time Control 1997}}, \bibinfo{pages}{109--114} (\bibinfo{year}{1998}).

\bibitem{hibbardAvoidingUnintendedAI2012}
\bibinfo{author}{Hibbard, B.}
\newblock \bibinfo{editor}{Bach, J.}, \bibinfo{editor}{Goertzel, B.} \& \bibinfo{editor}{Ikl{\'e}, M.} (eds) \emph{\bibinfo{title}{Avoiding {{Unintended AI Behaviors}}}}.
\newblock (eds \bibinfo{editor}{Bach, J.}, \bibinfo{editor}{Goertzel, B.} \& \bibinfo{editor}{Ikl{\'e}, M.}) \emph{\bibinfo{booktitle}{Artificial {{General Intelligence}}}}, Lecture {{Notes}} in {{Computer Science}}, \bibinfo{pages}{107--116} (\bibinfo{publisher}{{Springer}}, \bibinfo{address}{{Berlin, Heidelberg}}, \bibinfo{year}{2012}).

\bibitem{sezenerInferringHumanValues2015}
\bibinfo{author}{Sezener, {\relax CE}.}
\newblock \bibinfo{editor}{Bieger, J.}, \bibinfo{editor}{Goertzel, B.} \& \bibinfo{editor}{Potapov, A.} (eds) \emph{\bibinfo{title}{Inferring {{Human Values}} for {{Safe AGI Design}}}}.
\newblock (eds \bibinfo{editor}{Bieger, J.}, \bibinfo{editor}{Goertzel, B.} \& \bibinfo{editor}{Potapov, A.}) \emph{\bibinfo{booktitle}{Artificial General Intelligence ({{AGI}} 2015)}}, Vol. \bibinfo{volume}{9205}, \bibinfo{pages}{152--155} (\bibinfo{year}{2015}).

\bibitem{elmhamdiWhenNeuronsFail2017}
\bibinfo{author}{El~Mhamdi, E.} \& \bibinfo{author}{Guerraoui, R.}
\newblock \emph{\bibinfo{title}{When {{Neurons Fail}}}}, \bibinfo{pages}{1028--1037} (\bibinfo{year}{2017}).

\bibitem{freieslebenGeneralizationTheoryRobustness2023}
\bibinfo{author}{Freiesleben, T.} \& \bibinfo{author}{Grote, T.}
\newblock \bibinfo{title}{Beyond generalization: A theory of robustness in machine learning}.
\newblock \emph{\bibinfo{journal}{Synthese}} \textbf{\bibinfo{volume}{202}} (\bibinfo{year}{2023}).

\bibitem{sannemanTransparentValueAlignment2023}
\bibinfo{author}{Sanneman, L.} \& \bibinfo{author}{Shah, J.}
\newblock \emph{\bibinfo{title}{Transparent {{Value Alignment}}}}, \bibinfo{pages}{557--560} (\bibinfo{publisher}{{ACM}}, \bibinfo{address}{{Stockholm Sweden}}, \bibinfo{year}{2023}).

\bibitem{murphyLearningEffectiveInterpretable2012}
\bibinfo{author}{Murphy, B.}, \bibinfo{author}{Talukdar, P.} \& \bibinfo{author}{Mitchell, T.}
\newblock \bibinfo{editor}{Kay, M.} \& \bibinfo{editor}{Boitet, C.} (eds) \emph{\bibinfo{title}{Learning {{Effective}} and {{Interpretable Semantic Models}} using {{Non-Negative Sparse Embedding}}}}.
\newblock (eds \bibinfo{editor}{Kay, M.} \& \bibinfo{editor}{Boitet, C.}) \emph{\bibinfo{booktitle}{Proceedings of {{COLING}} 2012}}, \bibinfo{pages}{1933--1950} (\bibinfo{publisher}{{The COLING 2012 Organizing Committee}}, \bibinfo{address}{{Mumbai, India}}, \bibinfo{year}{2012}).

\bibitem{subramanianSPINESParseInterpretable2017}
\bibinfo{author}{Subramanian, A.}, \bibinfo{author}{Pruthi, D.}, \bibinfo{author}{Jhamtani, H.}, \bibinfo{author}{{Berg-Kirkpatrick}, T.} \& \bibinfo{author}{Hovy, E.}
\newblock \bibinfo{title}{{{SPINE}}: {{SParse Interpretable Neural Embeddings}}} (\bibinfo{year}{2017}).
\newblock \eprint{1711.08792}.

\bibitem{shahamUnderstandingAdversarialTraining2018}
\bibinfo{author}{Shaham, U.}, \bibinfo{author}{Yamada, Y.} \& \bibinfo{author}{Negahban, S.}
\newblock \bibinfo{title}{Understanding adversarial training: {{Increasing}} local stability of supervised models through robust optimization}.
\newblock \emph{\bibinfo{journal}{Neurocomputing}} \textbf{\bibinfo{volume}{307}}, \bibinfo{pages}{195--204} (\bibinfo{year}{2018}).

\bibitem{wuPUMAPerformanceUnchanged2022}
\bibinfo{author}{Wu, G.}, \bibinfo{author}{Hashemi, M.} \& \bibinfo{author}{Srinivasa, C.}
\newblock \bibinfo{title}{{{PUMA}}: {{Performance Unchanged Model Augmentation}} for {{Training Data Removal}}} (\bibinfo{year}{2022}).
\newblock \eprint{2203.00846}.

\bibitem{jingRobustExtremeLearning2020}
\bibinfo{author}{Jing, S.} \& \bibinfo{author}{Yang, L.}
\newblock \bibinfo{title}{A robust extreme learning machine framework for uncertain data classification}.
\newblock \emph{\bibinfo{journal}{Journal of Supercomputing}} \textbf{\bibinfo{volume}{76}}, \bibinfo{pages}{2390--2416} (\bibinfo{year}{2020}).

\bibitem{ganDualLearningBasedSafe2018}
\bibinfo{author}{Gan, {\relax HT}.}, \bibinfo{author}{Li, {\relax ZH}.}, \bibinfo{author}{Fan, {\relax YL}.} \& \bibinfo{author}{Luo, {\relax ZZ}.}
\newblock \bibinfo{title}{Dual {{Learning-Based Safe Semi-Supervised Learning}}}.
\newblock \emph{\bibinfo{journal}{IEEE Access}} \textbf{\bibinfo{volume}{6}}, \bibinfo{pages}{2615--2621} (\bibinfo{year}{2018}).

\bibitem{engstromAdversarialRobustnessPrior2019a}
\bibinfo{author}{Engstrom, L.} \emph{et~al.}
\newblock \bibinfo{title}{Adversarial {{Robustness}} as a {{Prior}} for {{Learned Representations}}}.
\newblock \emph{\bibinfo{journal}{{arXiv}}}  (\bibinfo{year}{2019}).

\bibitem{brophyMachineUnlearningRandom2021}
\bibinfo{author}{Brophy, J.} \& \bibinfo{author}{Lowd, D.}
\newblock \emph{\bibinfo{title}{Machine {{Unlearning}} for {{Random Forests}}}}, \bibinfo{pages}{1092--1104} (\bibinfo{publisher}{{PMLR}}, \bibinfo{year}{2021}).

\bibitem{chundawatZeroShotMachineUnlearning2023}
\bibinfo{author}{Chundawat, V.~S.}, \bibinfo{author}{Tarun, A.~K.}, \bibinfo{author}{Mandal, M.} \& \bibinfo{author}{Kankanhalli, M.}
\newblock \bibinfo{title}{Zero-{{Shot Machine Unlearning}}}.
\newblock \emph{\bibinfo{journal}{IEEE Transactions on Information Forensics and Security}} \textbf{\bibinfo{volume}{18}}, \bibinfo{pages}{2345--2354} (\bibinfo{year}{2023}).

\bibitem{chenATOMRobustifyingOutofdistribution2021}
\bibinfo{author}{Chen, J.}, \bibinfo{author}{Li, Y.}, \bibinfo{author}{Wu, X.}, \bibinfo{author}{Liang, Y.} \& \bibinfo{author}{Jha, S.}
\newblock \bibinfo{editor}{Oliver, N.}, \bibinfo{editor}{{P{\'e}rez-Cruz}, F.}, \bibinfo{editor}{Kramer, S.}, \bibinfo{editor}{Read, J.} \& \bibinfo{editor}{Lozano, J.~A.} (eds) \emph{\bibinfo{title}{{{ATOM}}: {{Robustifying Out-of-Distribution Detection Using Outlier Mining}}}}.
\newblock (eds \bibinfo{editor}{Oliver, N.}, \bibinfo{editor}{{P{\'e}rez-Cruz}, F.}, \bibinfo{editor}{Kramer, S.}, \bibinfo{editor}{Read, J.} \& \bibinfo{editor}{Lozano, J.~A.}) \emph{\bibinfo{booktitle}{Machine {{Learning}} and {{Knowledge Discovery}} in {{Databases}}. {{Research Track}}}}, Lecture {{Notes}} in {{Computer Science}}, \bibinfo{pages}{430--445} (\bibinfo{publisher}{{Springer International Publishing}}, \bibinfo{address}{{Cham}}, \bibinfo{year}{2021}).

\bibitem{lakkarajuIdentifyingUnknownUnknowns2016}
\bibinfo{author}{Lakkaraju, H.}, \bibinfo{author}{Kamar, E.}, \bibinfo{author}{Caruana, R.} \& \bibinfo{author}{Horvitz, E.}
\newblock \emph{\bibinfo{title}{Identifying unknown unknowns in the open world: Representations and policies for guided exploration}}, Vol.~\bibinfo{volume}{31} (\bibinfo{year}{2017}).

\bibitem{zhuoDeepUnsupervisedConvolutional2017}
\bibinfo{author}{Zhuo, {\relax JB}.}, \bibinfo{author}{Wang, {\relax SH}.}, \bibinfo{author}{Zhang, {\relax WG}.} \& \bibinfo{author}{Huang, {\relax QM}.}
\newblock \emph{\bibinfo{title}{Deep {{Unsupervised Convolutional Domain Adaptation}}}}, \bibinfo{pages}{261--269} (\bibinfo{year}{2017}).

\bibitem{bossensExplicitExploreExploit2023}
\bibinfo{author}{Bossens, {\relax DM}.} \& \bibinfo{author}{Bishop, N.}
\newblock \bibinfo{title}{Explicit {{Explore}}, {{Exploit}}, or {{Escape}} ({{E4}}): Near-optimal safety-constrained reinforcement learning in polynomial time}.
\newblock \emph{\bibinfo{journal}{Machine Learning}} \textbf{\bibinfo{volume}{112}}, \bibinfo{pages}{817--858} (\bibinfo{year}{2023}).

\bibitem{massianiSafeValueFunctions2023}
\bibinfo{author}{Massiani, {\relax PF}.}, \bibinfo{author}{Heim, S.}, \bibinfo{author}{Solowjow, F.} \& \bibinfo{author}{Trimpe, S.}
\newblock \bibinfo{title}{Safe {{Value Functions}}}.
\newblock \emph{\bibinfo{journal}{IEEE Transactions on Automatic Control}} \textbf{\bibinfo{volume}{68}}, \bibinfo{pages}{2743--2757} (\bibinfo{year}{2023}).

\bibitem{shiNearOptimalAlgorithmSafe2023}
\bibinfo{author}{Shi, M.}, \bibinfo{author}{Liang, Y.} \& \bibinfo{author}{Shroff, N.}
\newblock \emph{\bibinfo{title}{A {{Near-Optimal Algorithm}} for {{Safe Reinforcement Learning Under Instantaneous Hard Constraints}}}}, Vol. \bibinfo{volume}{202}, \bibinfo{pages}{31243--31268} (\bibinfo{year}{2023}).

\bibitem{huntVerifiablySafeExploration2021}
\bibinfo{author}{Hunt, N.} \emph{et~al.}
\newblock \emph{\bibinfo{title}{Verifiably {{Safe Exploration}} for {{End-to-End Reinforcement Learning}}}} (\bibinfo{year}{2021}).

\bibitem{maConservativeAdaptivePenalty2022}
\bibinfo{author}{Ma, Y.~J.}, \bibinfo{author}{Shen, A.}, \bibinfo{author}{Bastani, O.} \& \bibinfo{author}{Dinesh, J.}
\newblock \emph{\bibinfo{title}{Conservative and {{Adaptive Penalty}} for {{Model-Based Safe Reinforcement Learning}}}}, Vol.~\bibinfo{volume}{36}, \bibinfo{pages}{5404--5412} (\bibinfo{year}{2022}).

\bibitem{zwaneSafeTrajectorySampling2023}
\bibinfo{author}{Zwane, S.} \emph{et~al.}
\newblock \emph{\bibinfo{title}{Safe {{Trajectory Sampling}} in {{Model-Based Reinforcement Learning}}}}, Vol. \bibinfo{volume}{2023-August} (\bibinfo{year}{2023}).

\bibitem{fischerSamplingbasedInverseReinforcement2021}
\bibinfo{author}{Fischer, J.}, \bibinfo{author}{Eyberg, C.}, \bibinfo{author}{Werling, M.} \& \bibinfo{author}{Lauer, M.}
\newblock \emph{\bibinfo{title}{Sampling-based {{Inverse Reinforcement Learning Algorithms}} with {{Safety Constraints}}}}, \bibinfo{pages}{791--798} (\bibinfo{year}{2021}).

\bibitem{zhouRobustMeanFieldActorCritic2023}
\bibinfo{author}{Zhou, Z.}, \bibinfo{author}{Liu, G.} \& \bibinfo{author}{Zhou, M.}
\newblock \bibinfo{title}{A {{Robust Mean-Field Actor-Critic Reinforcement Learning Against Adversarial Perturbations}} on {{Agent States}}}.
\newblock \emph{\bibinfo{journal}{IEEE Transactions on Neural Networks and Learning Systems}} \bibinfo{pages}{1--12} (\bibinfo{year}{2023}).

\bibitem{bazzanAligningIndividualCollective2019}
\bibinfo{author}{Bazzan, {\relax ALC}.}
\newblock \bibinfo{title}{Aligning individual and collective welfare in complex socio-technical systems by combining metaheuristics and reinforcement learning}.
\newblock \emph{\bibinfo{journal}{Engineering Applications of Artificial Intelligence}} \textbf{\bibinfo{volume}{79}}, \bibinfo{pages}{23--33} (\bibinfo{year}{2019}).

\bibitem{christoffersenGetItWriting2023}
\bibinfo{author}{Christoffersen, P.~J.}, \bibinfo{author}{Haupt, A.~A.} \& \bibinfo{author}{{Hadfield-Menell}, D.}
\newblock \emph{\bibinfo{title}{Get {{It}} in {{Writing}}: {{Formal Contracts Mitigate Social Dilemmas}} in {{Multi-Agent RL}}}}, {{AAMAS}} '23, \bibinfo{pages}{448--456} (\bibinfo{publisher}{{International Foundation for Autonomous Agents and Multiagent Systems}}, \bibinfo{address}{{Richland, SC}}, \bibinfo{year}{2023}).

\bibitem{christianoDeepReinforcementLearning2017}
\bibinfo{author}{Christiano, P.~F.} \emph{et~al.}
\newblock \emph{\bibinfo{title}{Deep {{Reinforcement Learning}} from {{Human Preferences}}}}, Vol.~\bibinfo{volume}{30} (\bibinfo{publisher}{{Curran Associates, Inc.}}, \bibinfo{year}{2017}).

\bibitem{kaushikLearningDifferenceThat2020}
\bibinfo{author}{Kaushik, D.}, \bibinfo{author}{Hovy, E.} \& \bibinfo{author}{Lipton, Z.~C.}
\newblock \emph{\bibinfo{title}{Learning the {{Difference}} that {{Makes}} a {{Difference}} with {{Counterfactually-Augmented Data}}}} (\bibinfo{publisher}{{arXiv}}, \bibinfo{year}{2020}).
\newblock \eprint{1909.12434}.

\bibitem{liBridgingModelbasedSafety2022}
\bibinfo{author}{Li, {\relax ZY}.}, \bibinfo{author}{Zeng, J.}, \bibinfo{author}{Thirugnanam, A.} \& \bibinfo{author}{Sreenath, K.}
\newblock \bibinfo{title}{Bridging {{Model-based Safety}} and {{Model-free Reinforcement Learning}} through {{System Identification}} of {{Low Dimensional Linear Models}}}.
\newblock \emph{\bibinfo{journal}{Robotics: Science and System}} \textbf{\bibinfo{volume}{18}} (\bibinfo{year}{2022}).

\bibitem{zhuContactSafeReinforcementLearning2022}
\bibinfo{author}{Zhu, X.}, \bibinfo{author}{Kang, {\relax SC}.}, \bibinfo{author}{Chen, {\relax JY}.} \& \bibinfo{author}{{IEEE}}
\newblock \emph{\bibinfo{title}{A {{Contact-Safe Reinforcement Learning Framework}} for {{Contact-Rich Robot Manipulation}}}}, \bibinfo{pages}{2476--2482} (\bibinfo{year}{2022}).

\bibitem{terraSafetyVsEfficiency2020}
\bibinfo{author}{Terra, A.}, \bibinfo{author}{Riaz, H.}, \bibinfo{author}{Raizer, K.}, \bibinfo{author}{Hata, A.} \& \bibinfo{author}{Inam, R.}
\newblock \emph{\bibinfo{title}{Safety vs. {{Efficiency}}: {{AI-Based Risk Mitigation}} in {{Collaborative Robotics}}}}, \bibinfo{pages}{151--160} (\bibinfo{year}{2020}).

\bibitem{carliniEvaluatingRobustnessNeural2017}
\bibinfo{author}{Carlini, N.} \& \bibinfo{author}{Wagner, D.}
\newblock \emph{\bibinfo{title}{Towards {{Evaluating}} the {{Robustness}} of {{Neural Networks}}}}, \bibinfo{pages}{39--57} (\bibinfo{publisher}{{IEEE Computer Society}}, \bibinfo{year}{2017}).

\bibitem{nguyenDeepNeuralNetworks2015}
\bibinfo{author}{Nguyen, A.}, \bibinfo{author}{Yosinski, J.} \& \bibinfo{author}{Clune, J.}
\newblock \emph{\bibinfo{title}{Deep neural networks are easily fooled: {{High}} confidence predictions for unrecognizable images}}, \bibinfo{pages}{427--436} (\bibinfo{publisher}{{IEEE}}, \bibinfo{address}{{Boston, MA, USA}}, \bibinfo{year}{2015}).

\bibitem{kaufmannTestingRobustnessUnforeseen2019}
\bibinfo{author}{Kaufmann, M.} \emph{et~al.}
\newblock \bibinfo{title}{Testing {{Robustness Against Unforeseen Adversaries}}}.
\newblock \emph{\bibinfo{journal}{{arXiv}}}  (\bibinfo{year}{2019}).

\bibitem{hendrycksWhatWouldJiminy2021}
\bibinfo{author}{Hendrycks, D.} \emph{et~al.}
\newblock \emph{\bibinfo{title}{What {{Would Jiminy Cricket Do}}? {{Towards Agents That Behave Morally}}}} (\bibinfo{year}{2021}).

\bibitem{gardnerEvaluatingModelsLocal2020}
\bibinfo{author}{Gardner, M.} \emph{et~al.}
\newblock \bibinfo{editor}{Cohn, T.}, \bibinfo{editor}{He, Y.} \& \bibinfo{editor}{Liu, Y.} (eds) \emph{\bibinfo{title}{Evaluating {{Models}}' {{Local Decision Boundaries}} via {{Contrast Sets}}}}.
\newblock (eds \bibinfo{editor}{Cohn, T.}, \bibinfo{editor}{He, Y.} \& \bibinfo{editor}{Liu, Y.}) \emph{\bibinfo{booktitle}{Findings of the {{Association}} for {{Computational Linguistics}}: {{EMNLP}} 2020}}, \bibinfo{pages}{1307--1323} (\bibinfo{publisher}{{Association for Computational Linguistics}}, \bibinfo{address}{{Online}}, \bibinfo{year}{2020}).

\bibitem{spearsAssuringBehaviorAdaptive2006}
\bibinfo{author}{Spears, D.~F.}
\newblock \bibinfo{title}{ in \textit{Assuring the {{Behavior}} of {{Adaptive Agents}}}} (eds \bibinfo{editor}{Rouff, C.~A.}, \bibinfo{editor}{Hinchey, M.}, \bibinfo{editor}{Rash, J.}, \bibinfo{editor}{Truszkowski, W.} \& \bibinfo{editor}{{Gordon-Spears}, D.}) \emph{\bibinfo{booktitle}{Agent {{Technology}} from a {{Formal Perspective}}}} {{NASA Monographs}} in {{Systems}} and {{Software Engineering}}, \bibinfo{pages}{227--257} (\bibinfo{publisher}{{Springer}}, \bibinfo{address}{{London}}, \bibinfo{year}{2006}).

\bibitem{wozniakSafetyCasePattern2020}
\bibinfo{author}{Wozniak, E.}, \bibinfo{author}{C{\^a}rlan, C.}, \bibinfo{author}{{Acar-Celik}, E.} \& \bibinfo{author}{Putzer, H.}
\newblock \emph{\bibinfo{title}{A {{Safety Case Pattern}} for {{Systems}} with {{Machine Learning Components}}}}, Vol. \bibinfo{volume}{12235 LNCS}, \bibinfo{pages}{370--382} (\bibinfo{year}{2020}).

\bibitem{gholampourAdversarialRobustnessPhishing2023}
\bibinfo{author}{Gholampour, P.} \& \bibinfo{author}{Verma, R.}
\newblock \emph{\bibinfo{title}{Adversarial {{Robustness}} of {{Phishing Email Detection Models}}}}, \bibinfo{pages}{67--76} (\bibinfo{year}{2023}).

\bibitem{nandaProgressMeasuresGrokking2023}
\bibinfo{author}{Nanda, N.}, \bibinfo{author}{Chan, L.}, \bibinfo{author}{Lieberum, T.}, \bibinfo{author}{Smith, J.} \& \bibinfo{author}{Steinhardt, J.}
\newblock \bibinfo{title}{Progress measures for grokking via mechanistic interpretability}.
\newblock \emph{\bibinfo{journal}{{arXiv}}}  (\bibinfo{year}{2023}).

\bibitem{olssonIncontextLearningInduction2022}
\bibinfo{author}{Olsson, C.} \emph{et~al.}
\newblock \bibinfo{title}{In-context learning and induction heads}.
\newblock \emph{\bibinfo{journal}{{Transformer Circuits Thread}}}  (\bibinfo{year}{2022}).
\newblock \bibinfo{note}{{OpenAI}}.

\bibitem{karpathyVisualizingUnderstandingRecurrent2016}
\bibinfo{author}{Karpathy, A.}, \bibinfo{author}{Johnson, J.} \& \bibinfo{author}{{Fei-Fei}, L.}
\newblock \emph{\bibinfo{title}{Visualizing and {{Understanding Recurrent Networks}}}} (\bibinfo{publisher}{{arXiv}}, \bibinfo{year}{2016}).
\newblock \eprint{1506.02078}.

\bibitem{morcosImportanceSingleDirections2018}
\bibinfo{author}{Morcos, A.~S.}, \bibinfo{author}{Barrett, D. G.~T.}, \bibinfo{author}{Rabinowitz, N.~C.} \& \bibinfo{author}{Botvinick, M.}
\newblock \bibinfo{title}{On the importance of single directions for generalization}.
\newblock \emph{\bibinfo{journal}{{arXiv}}}  (\bibinfo{year}{2018}).

\bibitem{gyevnarCausalExplanationsSequential2024}
\bibinfo{author}{Gyevnar, B.}, \bibinfo{author}{Wang, C.}, \bibinfo{author}{Lucas, C.~G.}, \bibinfo{author}{Cohen, S.~B.} \& \bibinfo{author}{Albrecht, S.~V.}
\newblock \emph{\bibinfo{title}{Causal {{Explanations}} for {{Sequential Decision-Making}} in {{Multi-Agent Systems}}}}, {{AAMAS}} '24, \bibinfo{pages}{771--779} (\bibinfo{publisher}{{International Foundation for Autonomous Agents and Multiagent Systems}}, \bibinfo{address}{Richland, SC}, \bibinfo{year}{2024}).

\bibitem{okawaAutomaticExplorationProcess2020}
\bibinfo{author}{Okawa, Y.}, \bibinfo{author}{Sasaki, T.} \& \bibinfo{author}{Iwane, H.}
\newblock \emph{\bibinfo{title}{Automatic {{Exploration Process Adjustment}} for {{Safe Reinforcement Learning}} with {{Joint Chance Constraint Satisfaction}}}}, Vol.~\bibinfo{volume}{53}, \bibinfo{pages}{1588--1595} (\bibinfo{year}{2020}).

\bibitem{ganRiskDegreebasedSafe2016}
\bibinfo{author}{Gan, {\relax HT}.}, \bibinfo{author}{Luo, {\relax ZZ}.}, \bibinfo{author}{Meng, M.}, \bibinfo{author}{Ma, {\relax YL}.} \& \bibinfo{author}{She, {\relax QS}.}
\newblock \bibinfo{title}{A risk degree-based safe semi-supervised learning algorithm}.
\newblock \emph{\bibinfo{journal}{International Journal of Machine Learning and Cybernetics}} \textbf{\bibinfo{volume}{7}}, \bibinfo{pages}{85--94} (\bibinfo{year}{2016}).

\bibitem{zhangBarrierLyapunovFunctionBased2022}
\bibinfo{author}{Zhang, {\relax YX}.} \emph{et~al.}
\newblock \bibinfo{title}{Barrier {{Lyapunov Function-Based Safe Reinforcement Learning}} for {{Autonomous Vehicles With Optimized Backstepping}}}.
\newblock \emph{\bibinfo{journal}{IEEE Transactions on Neural Networks and Learning Systems}}  (\bibinfo{year}{2022}).

\bibitem{nicolaeAlgorithmicRobustnessSemisupervised2015}
\bibinfo{author}{Nicolae, M.-I.}, \bibinfo{author}{Sebban, M.}, \bibinfo{author}{Habrard, A.}, \bibinfo{author}{Gaussier, E.} \& \bibinfo{author}{Amini, M.-R.}
\newblock \emph{\bibinfo{title}{Algorithmic robustness for semi-supervised ({$\epsilon$}, {$\gamma$}, {$\tau$})-good metric learning}}, Vol. \bibinfo{volume}{9489}, \bibinfo{pages}{253--263} (\bibinfo{year}{2015}).

\bibitem{khanNoHarmNovel2023}
\bibinfo{author}{Khan, {\relax WU}.} \& \bibinfo{author}{Seto, E.}
\newblock \bibinfo{title}{A "{{Do No Harm}}" {{Novel Safety Checklist}} and {{Research Approach}} to {{Determine Whether}} to {{Launch}} an {{Artificial Intelligence-Based Medical Technology}}: {{Introducing}} the {{Biological-Psychological}}, {{Economic}}, and {{Social}} ({{BPES}}) {{Framework}}}.
\newblock \emph{\bibinfo{journal}{Journal of Medical Internet Research}} \textbf{\bibinfo{volume}{25}} (\bibinfo{year}{2023}).

\bibitem{schumegProposedVModelVerification2023}
\bibinfo{author}{Schumeg, B.}, \bibinfo{author}{Marotta, F.} \& \bibinfo{author}{Werner, B.}
\newblock \emph{\bibinfo{title}{Proposed {{V-Model}} for {{Verification}}, {{Validation}}, and {{Safety Activities}} for {{Artificial Intelligence}}}}, \bibinfo{pages}{61--66} (\bibinfo{year}{2023}).

\bibitem{kammConceptDynamicRobust2023}
\bibinfo{author}{Kamm, S.}, \bibinfo{author}{Sahlab, N.}, \bibinfo{author}{Jazdi, N.} \& \bibinfo{author}{Weyrich, M.}
\newblock \emph{\bibinfo{title}{A {{Concept}} for {{Dynamic}} and {{Robust Machine Learning}} with {{Context Modeling}} for {{Heterogeneous Manufacturing Data}}}}, Vol. \bibinfo{volume}{118}, \bibinfo{pages}{354--359} (\bibinfo{year}{2023}).

\bibitem{costaRobustLearningMethodology2023}
\bibinfo{author}{Costa, E.}, \bibinfo{author}{Rebello, C.}, \bibinfo{author}{Fontana, M.}, \bibinfo{author}{Schnitman, L.} \& \bibinfo{author}{Nogueira, I.}
\newblock \bibinfo{title}{A {{Robust Learning Methodology}} for {{Uncertainty-Aware Scientific Machine Learning Models}}}.
\newblock \emph{\bibinfo{journal}{Mathematics}} \textbf{\bibinfo{volume}{11}} (\bibinfo{year}{2023}).

\bibitem{aksjonovSafetyCriticalDecisionMakingControl2023}
\bibinfo{author}{Aksjonov, A.} \& \bibinfo{author}{Kyrki, V.}
\newblock \bibinfo{title}{A {{Safety-Critical Decision-Making}} and {{Control Framework Combining Machine-Learning-Based}} and {{Rule-Based Algorithms}}}.
\newblock \emph{\bibinfo{journal}{SAE International Journal of Vehicle Dynamics Stability And NVH}} \textbf{\bibinfo{volume}{7}}, \bibinfo{pages}{287--299} (\bibinfo{year}{2023}).

\bibitem{antikainenDeploymentModelExtend2021}
\bibinfo{author}{Antikainen, J.} \emph{et~al.}
\newblock \bibinfo{editor}{Yue, T.} \& \bibinfo{editor}{Mirakhorli, M.} (eds) \emph{\bibinfo{title}{A {{Deployment Model}} to {{Extend Ethically Aligned AI Implementation Method ECCOLA}}}}.
\newblock (eds \bibinfo{editor}{Yue, T.} \& \bibinfo{editor}{Mirakhorli, M.}) \emph{\bibinfo{booktitle}{29th IEEE International Requirements Engineering Conference Workshops ({{REW}} 2021)}}, \bibinfo{pages}{230--235} (\bibinfo{year}{2021}).

\bibitem{vakkuriECCOLAMethodImplementing2021}
\bibinfo{author}{Vakkuri, V.}, \bibinfo{author}{Kemell, {\relax KK}.}, \bibinfo{author}{Jantunen, M.}, \bibinfo{author}{Halme, E.} \& \bibinfo{author}{Abrahamsson, P.}
\newblock \bibinfo{title}{{{ECCOLA}} - {{A}} method for implementing ethically aligned {{AI}} systems}.
\newblock \emph{\bibinfo{journal}{Journal of Systems and Software}} \textbf{\bibinfo{volume}{182}} (\bibinfo{year}{2021}).

\bibitem{zhangFairRoverExplorativeModel2021}
\bibinfo{author}{Zhang, H.}, \bibinfo{author}{Shahbazi, N.}, \bibinfo{author}{Chu, X.} \& \bibinfo{author}{Asudeh, A.}
\newblock \emph{\bibinfo{title}{{{FairRover}}: {{Explorative}} model building for fair and responsible machine learning}} (\bibinfo{year}{2021}).

\bibitem{costonValidityPerspectiveEvaluating2023}
\bibinfo{author}{Coston, A.} \emph{et~al.}
\newblock \emph{\bibinfo{title}{A {{Validity Perspective}} on {{Evaluating}} the {{Justified Use}} of {{Data-driven Decision-making Algorithms}}}}, \bibinfo{pages}{690--704} (\bibinfo{year}{2023}).

\bibitem{gittensAdversarialPerspectiveAccuracy2022}
\bibinfo{author}{Gittens, A.}, \bibinfo{author}{Yener, B.} \& \bibinfo{author}{Yung, M.}
\newblock \bibinfo{title}{An {{Adversarial Perspective}} on {{Accuracy}}, {{Robustness}}, {{Fairness}}, and {{Privacy}}: {{Multilateral-Tradeoffs}} in {{Trustworthy ML}}}.
\newblock \emph{\bibinfo{journal}{IEEE Access}} \textbf{\bibinfo{volume}{10}}, \bibinfo{pages}{120850--120865} (\bibinfo{year}{2022}).

\bibitem{taylorAlignmentAdvancedMachine2016}
\bibinfo{author}{Taylor, J.}, \bibinfo{author}{Yudkowsky, E.}, \bibinfo{author}{LaVictoire, P.} \& \bibinfo{author}{Critch, A.}
\newblock \bibinfo{title}{Alignment for advanced machine learning systems}.
\newblock \emph{\bibinfo{journal}{Ethics of Artificial Intelligence}} \bibinfo{pages}{342--382} (\bibinfo{year}{2016}).

\bibitem{sotalaResponsesCatastrophicAGI2014}
\bibinfo{author}{Sotala, K.} \& \bibinfo{author}{Yampolskiy, R.~V.}
\newblock \bibinfo{title}{Responses to catastrophic {{AGI}} risk: A survey}.
\newblock \emph{\bibinfo{journal}{Physica Scripta}} \textbf{\bibinfo{volume}{90}}, \bibinfo{pages}{018001} (\bibinfo{year}{2014}).

\bibitem{johnsonMetacognitionArtificialIntelligence2022}
\bibinfo{author}{Johnson, B.}
\newblock \bibinfo{title}{Metacognition for artificial intelligence system safety-{{An}} approach to safe and desired behavior}.
\newblock \emph{\bibinfo{journal}{Safety Science}} \textbf{\bibinfo{volume}{151}} (\bibinfo{year}{2022}).

\bibitem{hatherallResponsibleAgencyAnswerability2022}
\bibinfo{author}{Hatherall, L.} \emph{et~al.}
\newblock \emph{\bibinfo{title}{Responsible {{Agency Through Answerability}}}} (\bibinfo{year}{2022}).

\bibitem{stahlEmbeddingResponsibilityIntelligent2023}
\bibinfo{author}{Stahl, {\relax BC}.}
\newblock \bibinfo{title}{Embedding responsibility in intelligent systems: From {{AI}} ethics to responsible {{AI}} ecosystems}.
\newblock \emph{\bibinfo{journal}{Scientific Reports}} \textbf{\bibinfo{volume}{13}} (\bibinfo{year}{2023}).

\bibitem{samarasingheCounterfactualLearningEnhancing2023}
\bibinfo{author}{Samarasinghe, D.}
\newblock \bibinfo{title}{Counterfactual learning in enhancing resilience in autonomous agent systems}.
\newblock \emph{\bibinfo{journal}{Frontiers in Artificial Intelligence}} \textbf{\bibinfo{volume}{6}} (\bibinfo{year}{2023}).

\bibitem{diemertSafetyIntegrityLevels2023}
\bibinfo{author}{Diemert, S.}, \bibinfo{author}{Millet, L.}, \bibinfo{author}{Groves, J.} \& \bibinfo{author}{Joyce, J.}
\newblock \emph{\bibinfo{title}{Safety {{Integrity Levels}} for {{Artificial Intelligence}}}}, Vol. \bibinfo{volume}{14182 LNCS}, \bibinfo{pages}{397--409} (\bibinfo{year}{2023}).

\bibitem{wangDataBanzhafRobust2023}
\bibinfo{author}{Wang, J.} \& \bibinfo{author}{Jia, R.}
\newblock \emph{\bibinfo{title}{Data {{Banzhaf}}: {{A Robust Data Valuation Framework}} for {{Machine Learning}}}}, Vol. \bibinfo{volume}{206}, \bibinfo{pages}{6388--6421} (\bibinfo{year}{2023}).

\bibitem{everittSelfModificationPolicyUtility2016}
\bibinfo{author}{Everitt, T.}, \bibinfo{author}{Filan, D.}, \bibinfo{author}{Daswani, M.} \& \bibinfo{author}{Hutter, M.}
\newblock \bibinfo{title}{Self-{{Modification}} of {{Policy}} and {{Utility Function}} in {{Rational Agents}}}.
\newblock \emph{\bibinfo{journal}{{arXiv}}}  (\bibinfo{year}{2016}).

\bibitem{badeaMoralityMachinesInterpretation2022}
\bibinfo{author}{Badea, C.} \& \bibinfo{author}{Artus, G.}
\newblock \bibinfo{editor}{Bramer, M.} \& \bibinfo{editor}{Stahl, F.} (eds) \emph{\bibinfo{title}{Morality, {{Machines}}, and the {{Interpretation Problem}}: {{A Value-based}}, {{Wittgensteinian Approach}} to {{Building Moral Agents}}}}.
\newblock (eds \bibinfo{editor}{Bramer, M.} \& \bibinfo{editor}{Stahl, F.}) \emph{\bibinfo{booktitle}{Artificial Intelligence, {{AI}} 2022}}, Vol.~\bibinfo{volume}{39}, \bibinfo{pages}{124--137} (\bibinfo{year}{2022}).

\bibitem{umbrelloBeneficialArtificialIntelligence2019}
\bibinfo{author}{Umbrello, S.}
\newblock \bibinfo{title}{Beneficial {{Artificial Intelligence Coordination}} by {{Means}} of a {{Value Sensitive Design Approach}}}.
\newblock \emph{\bibinfo{journal}{Big Data and Cognitive Computing}} \textbf{\bibinfo{volume}{3}} (\bibinfo{year}{2019}).

\bibitem{yampolskiySafetyEngineeringArtificial2012}
\bibinfo{author}{Yampolskiy, R.} \& \bibinfo{author}{Fox, J.}
\newblock \bibinfo{title}{Safety {{Engineering}} for {{Artificial General Intelligence}}}.
\newblock \emph{\bibinfo{journal}{Topoi}}  (\bibinfo{year}{2012}).

\bibitem{weldFirstLawRobotics2009}
\bibinfo{author}{Weld, D.} \& \bibinfo{author}{Etzioni, O.}
\newblock \bibinfo{editor}{Barley, M.} \emph{et~al.} (eds) \emph{\bibinfo{title}{The {{First Law}} of {{Robotics}}}}.
\newblock (eds \bibinfo{editor}{Barley, M.} \emph{et~al.}) \emph{\bibinfo{booktitle}{Safety and {{Security}} in {{Multiagent Systems}}}}, Lecture {{Notes}} in {{Computer Science}}, \bibinfo{pages}{90--100} (\bibinfo{publisher}{{Springer}}, \bibinfo{address}{{Berlin, Heidelberg}}, \bibinfo{year}{2009}).

\bibitem{matthiasResponsibilityGapAscribing2004}
\bibinfo{author}{Matthias, A.}
\newblock \bibinfo{title}{The responsibility gap: {{Ascribing}} responsibility for the actions of learning automata}.
\newblock \emph{\bibinfo{journal}{Ethics and Information Technology}} \textbf{\bibinfo{volume}{6}}, \bibinfo{pages}{175--183} (\bibinfo{year}{2004}).

\bibitem{farinaArtificialIntelligenceSystems2022}
\bibinfo{author}{Farina, L.}
\newblock \bibinfo{editor}{Muller, {\relax VC}.} (ed.) \emph{\bibinfo{title}{Artificial {{Intelligence Systems}}, {{Responsibility}} and {{Agential Self-Awareness}}}}.
\newblock (ed.\bibinfo{editor}{Muller, {\relax VC}.}) \emph{\bibinfo{booktitle}{Philosophy and Theory of Artificial Intelligence 2021}}, Vol.~\bibinfo{volume}{63}, \bibinfo{pages}{15--25} (\bibinfo{year}{2022}).

\bibitem{lee2017flight}
\bibinfo{author}{Lee, A.~T.}
\newblock \emph{\bibinfo{title}{Flight simulation: virtual environments in aviation}}  (\bibinfo{publisher}{Routledge}, \bibinfo{year}{2017}).

\bibitem{weidinger2023sociotechnical}
\bibinfo{author}{Weidinger, L.} \emph{et~al.}
\newblock \bibinfo{title}{Sociotechnical safety evaluation of generative ai systems}.
\newblock \emph{\bibinfo{journal}{arXiv preprint arXiv:2310.11986}}  (\bibinfo{year}{2023}).

\end{thebibliography}

\end{document}